# Relativistic Transport Approach for Nucleus-Nucleus Collisions from SIS to SPS Energies ⋆,⋆⋆

W. Ehehalt and W. Cassing

*Institut für Theoretische Physik, Universität Giessen, 35392 Giessen, Germany*

**Abstract**

We formulate a covariant transport approach for high energy nucleus-nucleus collisions where the real part of the nucleon selfenergies is fitted to nuclear matter properties which are evaluated on the basis of a NJL-type Lagrangian for the quark degrees of freedom. The parameters of the quark-model Lagrangian are fixed by the Gell-Mann, Oakes and Renner relation, the pion-nucleon $\Sigma$-term, the nucleon energy as well as the nuclear binding energy at saturation density $\rho_0$. We find the resulting scalar and vector selfenergies for nucleons to be well in line with either Dirac-Brueckner computations for $\rho \leq 2\rho_0$ or those from the phenomenological optical potential when accounting for a swelling of the nucleon at finite nuclear matter density. The meson-baryon interaction density is modelled to describe a decrease of the meson mass with baryon density. The imaginary part of the hadron selfenergies is determined by a string fragmentation model which accounts for the in-medium mass of hadrons in line with the 'chiral' dynamics employed. The applicability of the transport approach is demonstrated in comparison with experimental data from SIS to SPS energies. The enhancement of the $K^+/\pi^+$ ratio in A + A collisions compared to p + A reactions at AGS energies is reproduced within the 'chiral' dynamics. Furthermore, detailed predictions for the stopping in Pb + Pb collisions at 153 GeV/A are presented.

## 1 Introduction

The study of hot and dense nuclear matter by means of relativistic nucleus-nucleus collisions is the major aim of high energy heavy-ion physics. However, any conclusions about the nuclear properties at high temperature or baryon densities must rely on the comparison of experimental data with theoretical approaches based on nonequilibrium kinetic theory. Among these, the covariant RBUU approach [1–9], the QMD [10] or RQMD model [11] have been successfully used in the past. As a genuine feature of transport theories there are two essential ingredients: i.e. the *baryon (and meson) scalar and vector selfenergies* - which are neglected in a couple of approaches - as well as *in-medium elastic and inelastic cross sections* for all hadrons involved. Whereas in the low-energy regime these 'transport coefficients' can be calculated in the Dirac-Brueckner approach starting from the bare nucleon-nucleon interaction [12,13], this is no longer possible at high baryon density ($\rho_{\rm B} \geq$ 2-3$\rho_0$) and high temperature, since the number of independent hadronic degrees of freedom increases drastically and the nuclear system is expected to enter a phase where chiral symmetry is restored [14–17]. Such a phase transition is dynamically due to a change of the nonperturbative QCD vacuum at high temperature or baryon density and the chiral invariance of the interaction between quarks and gluons in the QCD Lagrangian. As a consequence the hadron selfenergies in the nuclear medium should change substantially especially close to the chiral phase transition and any transport theoretical study should include the generic properties of QCD that so far are known from nonperturbative computations on the lattice [18–21]. However, such nonperturbative calculations will not be possible for high baryon densities within the next years and we have to rely on suitable effective Lagrangians that lead to the same physical condensates and thermodynamic behaviour as the original QCD problem.

⋆ supported by DFG and GSI Darmstadt
⋆⋆ Part of the PhD thesis of W. Ehehalt

In this paper we aim at formulating a 'chiral' transport theory for the hadronic degrees of freedom, which in covariant notation formally can be written as a coupled set of transport equations for the phase-space distributions $f_{\rm h}(x,p)$ of hadron h [1–3,7,8], i.e.

$$\begin{aligned}
&\left\{\left(\Pi_\mu - \Pi_\nu \partial_\mu^p U_{\rm h}^\nu - M_{\rm h}^* \partial_\mu^p U_{\rm h}^{\rm S}\right)\partial_x^\mu + \left(\Pi_\nu \partial_\mu^x U_{\rm h}^\nu + M_{\rm h}^* \partial_\mu^x U_{\rm h}^{\rm S}\right)\partial_p^\mu\right\} f_{\rm h}(x,p) \\
&= \sum_{{\rm h}_2{\rm h}_3{\rm h}_4\ldots} \int d2d3d4\ldots [G^\dagger G]_{12\to34\ldots}\,\delta_\Gamma^4(\Pi+\Pi_2-\Pi_3-\Pi_4\ldots) \\
&\times \left\{ f_{{\rm h}_3}(x,p_3) f_{{\rm h}_4}(x,p_4) \bar{f}_{\rm h}(x,p) \bar{f}_{{\rm h}_2}(x,p_2) \right. \\
&\left. - f_{\rm h}(x,p) f_{{\rm h}_2}(x,p_2) \bar{f}_{{\rm h}_3}(x,p_3) \bar{f}_{{\rm h}_4}(x,p_4) \right\} \ldots \quad . \qquad (1)
\end{aligned}$$

In eq. (1) $U_{\rm h}^{\rm S}(x,p)$ and $U_{\rm h}^\mu(x,p)$ denote the real part of the scalar and vector hadron selfenergies, respectively, while $[G^+G]_{12\to34\ldots}\delta_\Gamma^4(\Pi+\Pi_2-\Pi_3-\Pi_4\ldots)$ is the 'transition rate' for the process $1+2\to3+4+\ldots$ which is taken to be on-shell in the semiclassical limit adopted[1]. The hadron quasi-particle

[1] The index $\Gamma$ at the $\delta$-function indicates that off-shell transitions of width $\Gamma$ should also be allowed. In the actual transport simulation, however, we use the on-shell limit $\Gamma = 0$.

properties in (1) are defined via the mass-shell constraint [7],

$$\delta(\Pi_\mu \Pi^\mu - M_{\rm h}^{*2}) \quad , \tag{2}$$

with effective masses and momenta given by

$$\begin{aligned} M_{\rm h}^*(x,p) &= M_{\rm h} + U_{\rm h}^{\rm S}(x,p) \\ \Pi^\mu(x,p) &= p^\mu - U_{\rm h}^\mu(x,p) \quad , \end{aligned} \tag{3}$$

while the phase-space factors

$$\bar{f}_{\rm h}(x,p) = 1 \pm f_{\rm h}(x,p) \tag{4}$$

are responsible for fermion Pauli-blocking or Bose enhancement, respectively, depending on the type of hadron in the final/initial channel. The dots in eq. (1) stand for further contributions to the collision term with more than two hadrons in the final/initial channels. The transport approach (1) is fully specified by $U_{\rm h}^{\rm S}(x,p)$ and $U_{\rm h}^\mu(x,p)$ ($\mu = 0,1,2,3$), which determine the mean-field propagation of the hadrons, and by the transition rates $G^\dagger G\, \delta^4(\ldots)$ in the collision term, that describe the scattering and hadron production/absorption rates.

The scalar and vector mean fields $U_{\rm h}^{\rm S}$ and $U_{\rm h}^\mu$ are conventionally determined in the mean-field limit from an effective hadronic Lagrangian density $\mathcal{L}_{\rm H}$ which is the sum of the Lagrangian density for the free fields $\mathcal{L}_{\rm h}^0$ and some interaction density $\mathcal{L}_{\rm H}^{\rm int}$, i.e.

$$\mathcal{L}_{\rm H} = \sum_{\rm h} \mathcal{L}_{\rm h}^0 + \mathcal{L}_{\rm H}^{\rm int} \quad . \tag{5}$$

The actual form of $\mathcal{L}_{\rm H}^{\rm int}$, however, is only known for more simple cases at low baryon density $\rho_{\rm B}$ and its general form at high $\rho_{\rm B}$ and for large relative momenta between the interacting hadrons - which is probed in nucleus-nucleus collisions up to 200 GeV/A - is essentially undetermined. This opens up a large parameter space for coupling constants $g_{\rm hh'}$, from factors at the vertices as well as respective powers in the hadron fields, which might lead to various density isomers in the nuclear equation of state or mesonic condensates, respectively.

In order to reduce this large parameter space and to incorporate aspects of chiral symmetry we here adopt the strategy to specify $\mathcal{L}_{\rm H}^{\rm int}$ for baryons via an effective Lagrangian for the underlying quark degrees of freedom $\mathcal{L}_{\rm q}$ for nuclear-matter phase-space configurations (see below). In fixing scalar and vector couplings as well as vertex cutoffs for the 'quarks' and by comparing the energy density for nuclear-matter configurations from $\mathcal{L}_{\rm q}$ with that of $\mathcal{L}_{\rm H}$ on the hadronic side, we can fit hadronic couplings and vertices in $\mathcal{L}_{\rm H}^{\rm int}$ even for high baryon densities and thus determine $U_{\rm h}^{\rm S}$ and $U_{\rm h}^{\mu}$ in the transport equation (1) in a less arbitrary way.

The 'hard' hadronic processes, on the other hand, which govern the r.h.s. of eq. (1), are modeled by the LUND string-fragmentation [22] which is known to describe inelastic hadronic reactions in a wide energy regime at zero baryon density. The medium modifications due to the hadron selfenergies, however, require to introduce some conserving approximations in line with the density dependent hadron masses. With the specifications of $U_{\rm h}^{\rm S}(x,p)$ and $U_{\rm h}^{\mu}(x,p)$ and the inelastic collision rates $G^{\dagger}G\delta(...)$ the transport approach (1), which will be denoted by **HSD** [2], is fully defined and can be confronted with experiment.

Our work thus is organized as follows: In Section 2 we will evaluate the nucleon selfenergies $U_{\rm N}^{\rm S}$, $U_{\rm N}^{\mu}$ on the basis of a NJL-type Lagrangian $\mathcal{L}_{\rm q}$ and model the meson-baryon interaction. In this respect we first fix the free parameters of our quark model interaction by the Gell-Mann, Oakes and Renner relation [23], the pion-nucleon $\Sigma$-term and the free 'nucleon' mass. We then extend our study to the computation of the energy density of nuclear matter configurations, discuss the necessary modification of the nucleon formfactor in the medium and present results for the nuclear equation of state at low and high baryon density. Scalar and vector selfenergies for the nucleons are obtained by fitting the coupling parameters in the covariant approach of Weber et al. [7] to the density dependence of the effective mass and the energy per nucleon from the NJL calculations. We compare our results to the low density, low energy Dirac phenomenology and discuss the extrapolations to high baryon density and large relative momenta. Furthermore, the meson-baryon interactions in $\mathcal{L}_{\rm H}^{\rm int}$ are described along the line of Kaplan and Nelson [24] essentially employing density-dependent meson masses. In Section 3 we specify the modifications of the familiar LUND string-fragmentation model [22], which models the imaginary part of the hadron selfenergies or collision rates, to include (as a first step) the modification of the hadron masses at finite baryon density. In Section 4 we apply our transport approach to nucleus-nucleus collisions from SIS to SPS energies and test its applicability in comparison with experimental data. As a first test for the partial restoration of chiral symmetry in heavy-ion collisions we compute the $K^+/\pi^+$ ratio for systems at AGS energies and compare to the available data. Section 5, finally, is devoted to a summary and discussion of open problems.

[2] **H**adron-**S**tring-**D**ynamics

## 2 Hadron selfenergies

In this section we specify the evaluation of the mean fields $U_{\rm h}^{\rm S}$ and $U_{\rm h}^{\mu}$ that enter the l.h.s. of the transport equation (1) for the mean-field propagation. In order to reduce the parameter space and to obtain extrapolations for $U_{\rm h}^{\rm S}$ and $U_{\rm h}^{\mu}$ at high baryon density, we use an effective Lagrangian density $\mathcal{L}_{\rm q}$ for quarks that is compatible with the approximate chiral invariance of the QCD Lagrangian. We thus first fix the effective Lagrangian $\mathcal{L}_{\rm q}$ for the quark degrees of freedom on the mean-field (one loop) level for low energy QCD problems, where the gluon fields $A_{\mu}^{a}(x)$ are supposed to be integrated out. The effective interaction determined in this way should not be used in further perturbation theory (e.g. for scattering or transition rates) since it is assumed to be the result of an infinite resummation of interaction diagrams. In a second step we then extract the real part of nucleon selfenergies from quark configurations that describe nuclear matter at finite density.

### *2.1 The effective quark Lagrangian*

The underlying idea of an effective 4-point interaction for quarks has already been discussed e.g. by Vogl and Weise in ref. [25]. Since the fundamental currents in QCD are color currents, i.e. $J_{\mu}^{a} = \bar{\psi}_{\rm q}\gamma_{\mu}t^{a}\psi_{\rm q}$, an elementary color current interaction with a universal coupling $G_{\rm C}$ is expected to be dominant. An effective Lagrangian for $\delta(x_1 - x_2)$-like quark interactions thus reads

$$\mathcal{L}_{\rm q}(x) = \bar{\psi}_{\rm q}(\mathrm{i}\gamma^{\mu}\partial_{\mu} - \hat{m}_0)\psi_{\rm q} - G_{\rm C}^2 \sum_{a=1}^{8} \left(\bar{\psi}_{\rm q}\gamma_{\mu}t^{a}\psi_{\rm q}\right)^2 , \tag{6}$$

where $t^a (a = 1, ..., 8)$ are the $SU(3)_{\rm color}$ matrices with $tr(t^a t^b) = \delta_{ab}/2, \hat{m}_0$ a diagonal mass matrix in flavor space, i.e. $\hat{m}_0 = diag(m_{\rm u}^0, m_{\rm d}^0, m_{\rm s}^0)$ and $\bar{\psi}_{\rm q} = (\bar{\rm u}, \bar{\rm d}, \bar{\rm s})$ is the quark spinor in case of $SU(3)_{\rm flavor}$. The color-current interaction is invariant under chiral transformations or $SU(3)_{\rm flavor}$ rotations. The Lagrangian (6), however, in its present form is not yet well suited for the formulation of quark dynamics on the mean-field level because antisymmetrization generates a further mixing of color, flavor and Dirac indices. It is thus more convenient to introduce a Fierz transformation, i.e. to antisymmetrize the 4-point interaction to proceed with further computations on the Hartree level. The Fierz transform then generates color singlet as well as color octet terms, i.e. [25,26]

$$\mathcal{L}_{\rm q}(x) = \bar{\psi}_{\rm q}(\mathrm{i}\gamma^{\mu}\partial_{\mu} - \hat{m}_0)\psi_{\rm q} + G_{\rm S}^2 \sum_{i=0}^{8} \left\{ \left(\bar{\psi}_{\rm q}\frac{\lambda_i}{2}\psi_{\rm q}\right)^2 + \left(\bar{\psi}_{\rm q}\mathrm{i}\gamma_5\frac{\lambda_i}{2}\psi_{\rm q}\right)^2 \right\}$$

$$
-G_{\rm V}^2 \sum_{i=0}^{8} \left\{ \left( \bar\psi_{\rm q}\gamma_\mu \frac{\lambda_i}{2}\psi_{\rm q}\right)^2 + \left( \bar\psi_{\rm q}\gamma_5\gamma^\mu \frac{\lambda_i}{2}\psi_{\rm q}\right)^2 \right\}
$$

$$
-G_{\rm C}^2 \sum_{a=1}^{8} \left( \bar\psi_{\rm q}\gamma_\mu t^a \psi_{\rm q}\right)^2 - \frac{2}{3} G_{\rm C}^2 \sum_{a=1}^{8}\sum_{i=0}^{8} \left\{ \left( \bar\psi_{\rm q} \frac{\lambda^i}{2} t^a \psi_{\rm q}\right)^2 + \left( \bar\psi_{\rm q} {\rm i}\gamma_5 \frac{\lambda^i}{2} t^a \psi_{\rm q}\right)^2 \right\}
$$

$$
+\frac{1}{3} G_{\rm C}^2 \sum_{a=1}^{8}\sum_{i=0}^{8} \left\{ \left( \bar\psi_{\rm q}\gamma_\mu \frac{\lambda^i}{2} t^a \psi_{\rm q}\right)^2 + \left( \bar\psi_{\rm q}\gamma_\mu\gamma_5 \frac{\lambda^i}{2} t^a \psi_{\rm q}\right)^2 \right\} , \qquad (7)
$$

where $G_{\rm S}^2 = 2G_{\rm V}^2 = \frac{8}{9}G_{\rm C}^2$. In (7) the matrices $\lambda^i (i = 1,..,8)$ stand for the $SU(3)_{\rm flavor}$ degrees of freedom with $tr(\lambda_i\lambda_j) = 2\delta_{ij}$ while $\lambda^0$ is given by $\lambda^0 = \sqrt{\frac{2}{3}} I_3$ with $I_3$ denoting the 3x3 unitary matrix in flavor space[3]. The Lagrangian (7) in its color-singlet version has been the starting point for RPA-type calculations for the bosonic excitations of the nonperturbative QCD vacuum, i.e. the mesonic degrees of freedom [25–27]. Similar Lagrangian densities have also been exploited by a variety of authors [28–36] following an early suggestion by Nambu and Jona-Lasinio (NJL) [37].

In our present study we will discard the mesonic (RPA-type) sector and concentrate on the determination of a static effective quark-quark interaction by nucleon properties as well as nuclear matter related quantities. Similar concepts have been proposed by Guichon [38] and Saito and Thomas [39] based on bag-model wavefunctions. Here we start with a slightly different concept by determining the quark wavefunctions for the nucleon from the experimental data for the proton electromagnetic formfactor. In this way we intend to circumvent the problem of absolute confinement which cannot be dealt with properly using only a color neutral mean-field approach of the NJL-type. Since we are only interested in energy densities for *given* quark configurations, the resulting Lagrangian should *not* be used for dynamical studies such as the RPA response (mesonic sector). Furthermore, it is not expected that the respective soliton solution of (7) for a nucleon presents a dynamically stable object of a shape consistent with the experimental proton formfactor.

### *2.2 The isospin symmetric nucleon system*

In this Subsection we concentrate on vacuum as well as nucleon properties, where the nucleons are assumed to be represented by 3 additional valence quarks with a fixed phase-space distribution on top of the (truncated) Dirac sea with a formfactor in line with the experimental data. The singlet terms of the Lagrangian (7) in the mean-field limit - performing the sum over the flavor matrix elements - then leads to the following Lagrangian for $\bar{\psi}_k = (\bar{\mathrm{u}}, \bar{\mathrm{d}}, \bar{\mathrm{s}})$,

$$\begin{aligned} \mathcal{L}_{\mathrm{q}}(x) = \sum_{k=\mathrm{u,d,s}} & \left\{ \bar{\psi}_k \left( \mathrm{i}\gamma^\mu \partial_\mu - m_k^0 \right) \psi_k \right. \\ & + \frac{G_{\mathrm{S}}^2}{2} \left\{ \left( \bar{\psi}_k \psi_k \right)^2 + \left( \bar{\psi}_k \mathrm{i}\gamma_5 \psi_k \right)^2 \right\} \\ & \left. - \frac{G_{\mathrm{V}}^2}{2} \left\{ \left( \bar{\psi}_k \gamma_\mu \psi_k \right)^2 + \left( \bar{\psi}_k \gamma_5 \gamma^\mu \psi_k \right)^2 \right\} \right\}, \end{aligned} \tag{8}$$

where the couplings $G_{\mathrm{S}}^2$ and $G_{\mathrm{V}}^2$ are now considered as free parameters. For the systems of positive parity, which are of interest in our present work, also the pseudoscalar and pseudovector terms vanish in the Hartree limit such that we are left with the scalar and vector term, only. This is quite similar to the $\sigma$-$\omega$ model [40,41] in the nuclear physics context. The hamiltonian density then is given by

$$\begin{aligned} \mathcal{H}(x) = \sum_{k=\mathrm{u,d,s}} & \left\{ \bar{\psi}_k \left( -\mathrm{i}\gamma^i \partial_i + m_k^0 - G_{\mathrm{S}}^2 \langle \bar{\psi}_k \psi_k \rangle \right) \psi_k \right. \\ & \left. + \frac{G_{\mathrm{S}}^2}{2} \langle \bar{\psi}_k \psi_k \rangle^2 + \frac{G_{\mathrm{V}}^2}{2} \langle \bar{\psi}_k \gamma_\mu \psi_k \rangle^2 \right\}, \end{aligned} \tag{9}$$

which leads to the gap equations for the effective masses $m_k$, i.e.

[3] The most general four-point interaction compatible with QCD symmetries starts from combinations of all possible vector and axial currents. Therefore, in general, there is no strict relationship between $G_{\rm S}$ and $G_{\rm V}$.

$$m_k = m_k^0 - G_{\mathrm{S}}^2 \langle \bar{\psi}_k \psi_k \rangle. \tag{10}$$

Since the problem (9) decouples in the flavor degrees of freedom we will consider in the following only u-quarks assuming $m_{\mathrm{u}}^0 = m_{\mathrm{d}}^0$ and neglect a possible strangeness content of the nucleon furtheron.

For the nonperturbative vacuum we then end up with the gap equation in phase space for the effective quark mass $m_{\mathrm{u}}$ of u or d quarks:

$$m_{\mathrm{u}} = m_{\mathrm{u}}^0 + G_{\mathrm{S}}^2 \frac{g}{(2\pi)^3} \int d^3p \frac{m_{\mathrm{u}}}{\sqrt{p^2 + m_{\mathrm{u}}^2}} \; \Theta(\Lambda_{\mathrm{S}} - \mid \mathbf{p} \mid) = m_V, \tag{11}$$

where we have introduced a spatial cutoff parameter $\Lambda_{\mathrm{S}}$ to regularize the divergent integral over the Dirac sea. Alternatively, one might also introduce covariant cutoff schemes as in [25,34], but for reasons to be discussed below in context of eq. (13) we prefer to use the scheme (11), since we are basically interested in quark configurations with a well defined rest frame. In eq. (11) the factor $g = 6$ arizes from the trace over color and spin in eq. (10). The gap equation (11) then leads to a constituent quark mass $m_{\mathrm{u}} > m_{\mathrm{u}}^0$ in the nonperturbative vacuum.

The coupling constant $G_{\rm S}$ together with the cutoff parameter $\Lambda_{\rm S}$ now can be determined via the Gell-Mann, Oakes and Renner relation [23] assuming $\langle \bar{\rm u}{\rm u}\rangle = \langle \bar{\rm d}{\rm d}\rangle$,

$$m_\pi^2 \; f_\pi^2 = -(m_{\rm u}^0 + m_{\rm d}^0) \; \langle \bar{\rm u}{\rm u}\rangle \; , \tag{12}$$

where $f_\pi = 93.3$ MeV is the pion decay constant, $m_\pi$ the physical pion mass and $\langle \bar{\rm u}{\rm u}\rangle$ the scalar condensate (for u or d quarks in the vakuum). Choosing $m_{\rm u}^0 = 7$ MeV as an average value of the light quark mass the quark condensate then amounts to $\langle \bar{\rm u}{\rm u}\rangle^{1/3} \approx -230$ MeV; a value which is achieved by choosing a cutoff $\Lambda_{\rm S} \approx 0.59$ GeV and $G_{\rm S} \approx 4.95$ GeV$^{-1}$ in (11).

In the presense of additional localized light valence quarks on top of the Dirac sea the gap equation (11) modifies locally to

$$m_{\rm u}({\bf r}) = m_{\rm u}^0 - G_{\rm S}^2 \frac{g}{(2\pi)^3} \int d^3p \, \frac{m_{\rm u}({\bf r})}{\sqrt{p^2 + m_{\rm u}({\bf r})^2}} \; f_{\rm u}({\bf r},{\bf p})$$
$$+ G_{\rm S}^2 \frac{g}{(2\pi)^3} \int d^3p \, \frac{m_{\rm u}({\bf r})}{\sqrt{p^2 + m_{\rm u}({\bf r})^2}} \; \Theta(\Lambda_{\rm S} - \mid {\bf p} \mid), \tag{13}$$

where $f_{\rm u}({\bf r},{\bf p})$ denotes the phase-space distribution of a single u-quark which has to determined in a model dependent way.

In a fully dynamical theory on the mean-field level $f_{\rm u}({\bf r},{\bf p})$ should result from the solution of the Dirac equation

$$\{ -{\rm i}\gamma^i \partial_i + m_k^0 - G_{\rm S}^2 \rho_{\rm S}({\bf r}) + \gamma^0 \rho_{\rm V}({\bf r}) \} \psi_k({\bf r}) = \gamma^0 E_k \psi_k({\bf r}) \tag{14}$$

with

$$\rho_{\rm S}({\bf r}) = \langle \bar{\psi}_k({\bf r}) \psi_k({\bf r}) \rangle \; ,$$
$$\rho_{\rm V}({\bf r}) = \langle \psi_k^\dagger({\bf r}) \psi_k({\bf r}) \rangle \; , \tag{15}$$

and subsequent Wigner-transformation of $\sum_k \psi_k^\dagger({\bf r} - {\bf s}/2) \psi_k({\bf r} + {\bf s}/2)$. However, since we do not aim at a dynamical theory for the nucleon - due to the lack of confinement in $\mathcal{L}_{\rm q}(x)$ (8) - and we are only interested in the total energy of well defined quark configurations, we fix $f_{\rm u}({\bf r},{\bf p})$ (from outside) by the experimental electromagnetic formfactor of the proton which is well represented by a dipole approximation up to momentum transfers $Q^2 \approx 25$ GeV$^2$/c$^2$ [42]. This implies that the quark charge distribution (of a proton) is of the exponential form [43]

$$\langle \psi_{\rm q}^\dagger({\bf r}) \psi_{\rm q}({\bf r}) \rangle \approx N_0 \; exp(- \mid {\bf r} \mid / b_0) = \rho_{\rm q}({\bf r}), \tag{16}$$

where $\mathbf{r}$ is given in fm, $b_0 = 0.25$ fm and $N_0 = (8\pi b_0^3)^{-1}$ provides normalization to 1. Considering now a nucleon state averaged over spin and isospin, i.e. a mixture of proton, neutron and $\Delta'$s of average mass $M_{\rm N} \approx 1.085$ GeV, we obtain for the u-quark density

$$\rho_{\rm u}(\mathbf{r}) \approx \frac{3}{2} N_0 \; exp(- \mid \mathbf{r} \mid / b_0), \tag{17}$$

where the factor 3/2 reflects the average u-quark content of the states considered. In the local density approximation the phase-space distribution for u-quarks then is given by

$$f_{\rm u}(\mathbf{r}, \mathbf{p}) = \Theta(p_{\rm F}(\mathbf{r}) - \mid \mathbf{p} \mid) \tag{18}$$

with the local Fermi momentum

$$p_{\rm F}(\mathbf{r}) = (6/g \; \pi^2)^{1/3} \rho_{\rm u}(\mathbf{r})^{1/3}. \tag{19}$$

This approximation has been quite successfully applied in the nuclear physics context [1,2] and also been adopted in [44,45] for quark oriented models. It is a legitimate approximation for the quark phase-space distribution as long as one is interested in expectation values like the total energy, only.

Inserting $f_{\rm u}(\mathbf{r}, \mathbf{p})$ (18) with (17) and (19) in the gap equation (13) we can compute the effective quark mass $m_{\rm u}(\mathbf{r})$ for the 'nucleon' described above. The resulting coordinate-space dependence of $m_{\rm u}(\mathbf{r})$ (full line) for the 'nucleon' is shown in Fig. 1 together with the u-quark density $\langle {\rm u}^\dagger(r) {\rm u}(r) \rangle = \rho_{\rm u}(r)$ (dashed line). In the interior of the 'nucleon' the effective quark mass drops to about $m_{\rm u}^0 = 7$ MeV and thus the quark scalar selfenergy $U_{\rm S}^{\rm q}$ to zero.

Whereas the scalar sector now is fixed by the gap equation (13) for arbitrary quark phase-space distributions $f_{\rm u}(\mathbf{r}, \mathbf{p})$ - that are at rest within the frame of reference considered here - the local vector quark interaction is modified in order to allow for an explicite momentum dependence. We note that nonlocal generalizations of the NJL Lagrangian have been suggested by Bowler and Birse [46]. We adopt a similar concept and assume that the vector interaction in (8) is mediated by massive color neutral (vector) gluons which implies to modify the couplings

$$G_{\rm V} \to G_{\rm V} \frac{\Lambda_{\rm V}^2}{\Lambda_{\rm V}^2 + \mathbf{q}^2} \quad , \tag{20}$$

where $\Lambda_{\rm V} \approx 1.2 - 1.5$ GeV is a vector cutoff and $\mathbf{q}$ denotes the momentum transfer in the quark-quark interaction. This strategy is similar to that used in effective meson-exchange interactions for hadron-hadron scattering [47].

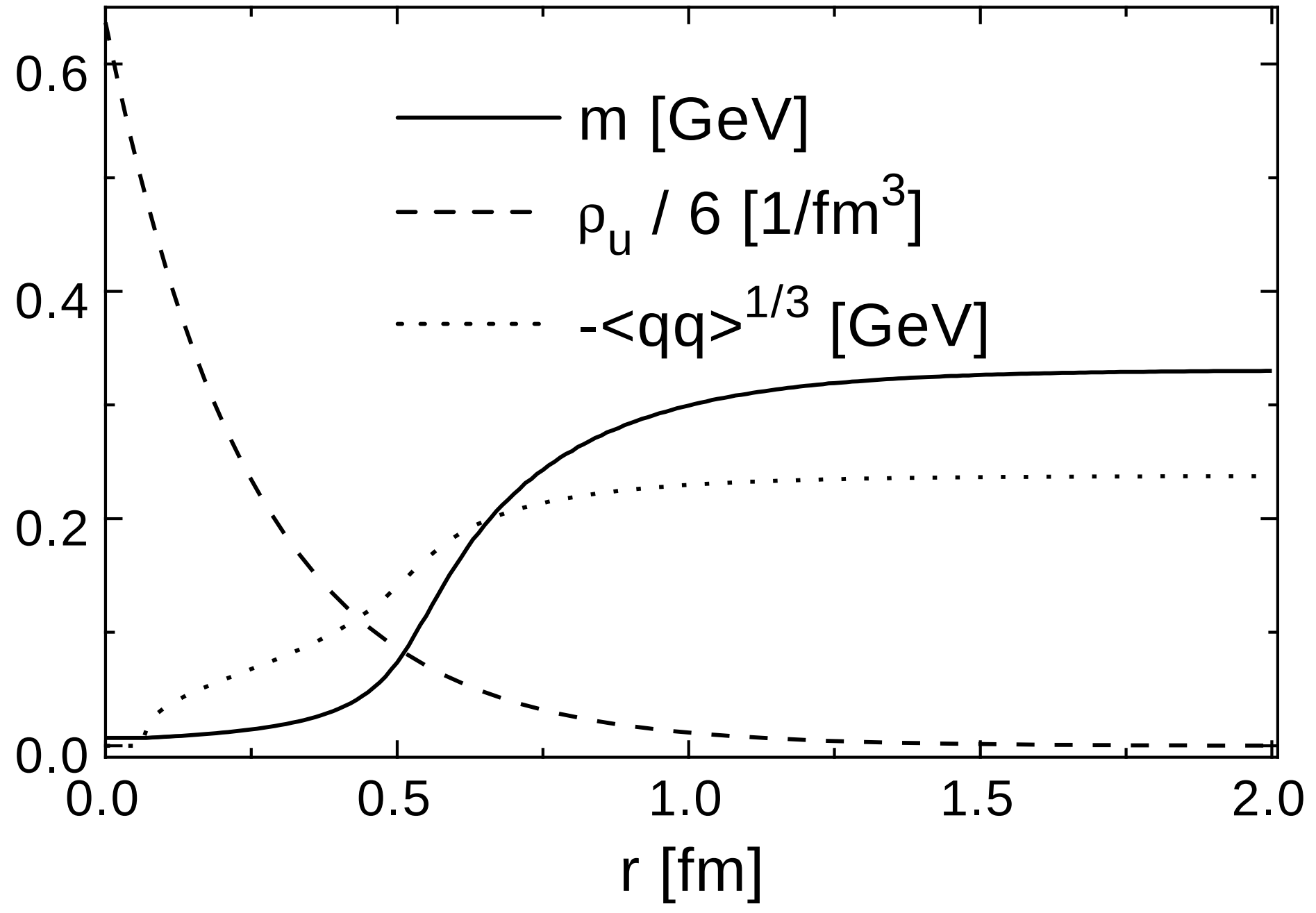

Fig. 1. Effective quark mass $m = m_{\rm u}(r)$ (full line), quark density $\rho_{\rm u}(r)$ (dashed line) and scalar condensate $- < \bar{q}q >^{1/3}$ (dotted line) as a function of the radial distance $r$ from the center of the 'nucleon'.

The energy density $T^{00}(\mathbf{r})$ in phase-space representation thus reads (including a factor of 2 from the summation over u and d quarks)

$$
\begin{aligned}
T^{00}(\mathbf{r}) = & 2g \int \frac{d^3p}{(2\pi)^3} \sqrt{p^2 + m_{\rm u}(\mathbf{r})^2}\, f_{\rm u}(\mathbf{r},\mathbf{p}) \\
& -2g \int \frac{d^3p}{(2\pi)^3} \sqrt{p^2 + m_{\rm u}(\mathbf{r})^2}\ \Theta(\Lambda_{\rm S} - \mid \mathbf{p} \mid) \\
& +2 \left\{ \frac{1}{2} G_{\rm S}^2 \rho_{\rm S}(\mathbf{r})^2 + \frac{1}{2} G_{\rm V}^2 \frac{g^2}{(2\pi)^6} \int d^3p_1 d^3p_2 f_{\rm u}(\mathbf{r},\mathbf{p}_1) \frac{\Lambda_{\rm V}^2}{\Lambda_{\rm V}^2 + (\mathbf{p_1} - \mathbf{p_2})^2} f_{\rm u}(\mathbf{r},\mathbf{p}_2) \right\} \\
& -E_{\rm vac},
\end{aligned}
\tag{21}
$$

where the vacuum contribution

$$
E_{\rm vac} = -2 \left\{ g \int \frac{d^3p}{(2\pi)^3} \sqrt{p^2 + m_{\rm u}^2}\ \Theta(\Lambda_{\rm S} - \mid \mathbf{p} \mid) + \frac{1}{2} G_{\rm S}^2 \rho_{\rm S0}^2 \right\}
\tag{22}
$$

has been subtracted. In eq. (21) $\rho_{\rm S} = (m_{\rm u} - m^0)/G_{\rm S}^2$ is the scalar quark density and the vector quark density is $\rho_{\rm V} = \frac{g}{(2\pi)^3} \int d^3p\ f_{\rm u}(\mathbf{r},\mathbf{p})$. The total energy $\langle H \rangle$ of a quark configuration described by $f_{\rm u}(\mathbf{r},\mathbf{p})$ then is obtained by integrating $\int d^3r\ T^{00}(\mathbf{r})$.

The average nucleon energy to be fixed in our case corresponds to 1.085 GeV, which is the average of the nucleon and the $\Delta$ mass. Since in eq. (21) for $T^{00}(\mathbf{r})$ all quantities are determined except the quark vector coupling $G_{\rm V}$ and cutoff $\Lambda_{\rm V}$, the vector coupling (for fixed $\Lambda_{\rm V} \approx 1.5$ GeV) is well determined by the total energy of the quark configuration. Our fit provides $G_{\rm V} = 4.2$ GeV$^{-1}$ using $f_{\rm u}(\mathbf{r},\mathbf{p}) = \Theta(p_{\rm F}(\mathbf{r}) - \mid \mathbf{p} \mid)$ with $p_{\rm F}(\mathbf{r})$ from (19). The pion-nucleon $\Sigma$-term, defined by the following matrix element with the nucleon state,

$$\Sigma_{\pi {\rm N}} = \frac{1}{2}(m_{\rm u}^0 + m_{\rm d}^0)\ \langle {\rm N}|\bar{\rm u}{\rm u} + \bar{\rm d}{\rm d}|{\rm N}\rangle\ , \tag{23}$$

within the parameters stated above leads to $\Sigma_{\pi{\rm N}} \approx 47$ MeV, which is well in line with the value extracted from pion-nucleon s-wave scattering of $45 \pm 7$ MeV from [48]. This will be of significant importance for the scalar nucleon selfenergy later on.

We stress again that the coupling parameters $G_{\rm S}$, $G_{\rm V}$ and cutoffs $\Lambda_{\rm S}$, $\Lambda_{\rm V}$ only apply for the semiclassical static quark configurations discussed so far and should not be considered as appropriate for a fully dynamical theory on the basis of the Lagrangian (8). In fact, the vector coupling $G_{\rm V}$ is larger than in refs. [25–27] where the mesonic sector has been explored. As a consequence we can only attempt to describe 'nucleons' at finite baryon density and have to discard mesonic degrees of freedom.

### *2.3 Symmetric nuclear matter*

In order to evaluate the energy density for symmetric nuclear matter configurations we have to introduce in addition to $f_{\rm q} = f_{\rm u}(\mathbf{r},\mathbf{p})$ a phase-space distribution for the nucleons or 'localized' quark states $f_{\rm N}$. Denoting by $(\mathbf{r}_{\rm N}, \mathbf{p}_{\rm N})$ the position and momentum of a nucleon, the corresponding quark phase-space distribution $f_{\rm q}(\mathbf{r},\mathbf{p})_{\mathbf{r}_{\rm N},\mathbf{p}_{\rm N}}$ is obtained from a translation of the center of $f_{\rm q}$ by $\mathbf{r}_{\rm N}$ and a proper Lorentz transformation by $\beta_{\rm N} = \mathbf{p}/\sqrt{p^2 + m_{\rm N}^2}$ in phase space, i.e. a contraction of $f_{\rm q}$ by $\gamma_{\rm N}^{-1} = \sqrt{1-\beta_{\rm N}^2}$ in coordinate space and dilation in momentum space by $\gamma_{\rm N}$, which keeps the individual phase space integral invariant.

Before going over to the nuclear matter problem we first consider two-nucleon configurations for 'frozen' nucleon quark distributions [4] in the nucleon-nucleon center-of-mass system (c.m.s.). As an example the local quark phase-space distribution $f_{\rm u}(\mathbf{r},\mathbf{p})$ - as met in the overlap regime of two colliding quark states - is depicted in Fig. 2 as a function of $p_x$ and $p_z$ for $p_y = 0$. It's macroscopic pa-rameters are given by the relative momenta $P_1, P_2$ of the 'quark wave functions' with respect to the nuclear matter rest frame and the individual Fermi momenta $p_{F1}(\mathbf{r}), p_{F2}(\mathbf{r})$ that are determined by the individual densities at space position $\mathbf{r}$ by $p_{Fi}(\mathbf{r}) = (6/g\pi^2\rho_u^i(\mathbf{r}))^{1/3}$ as before. The Dirac sea contribution at rest is indicated by the sphere with radius $\Lambda_S$.

[4] This is denoted as the 'sudden' approximation in the nuclear physics context.

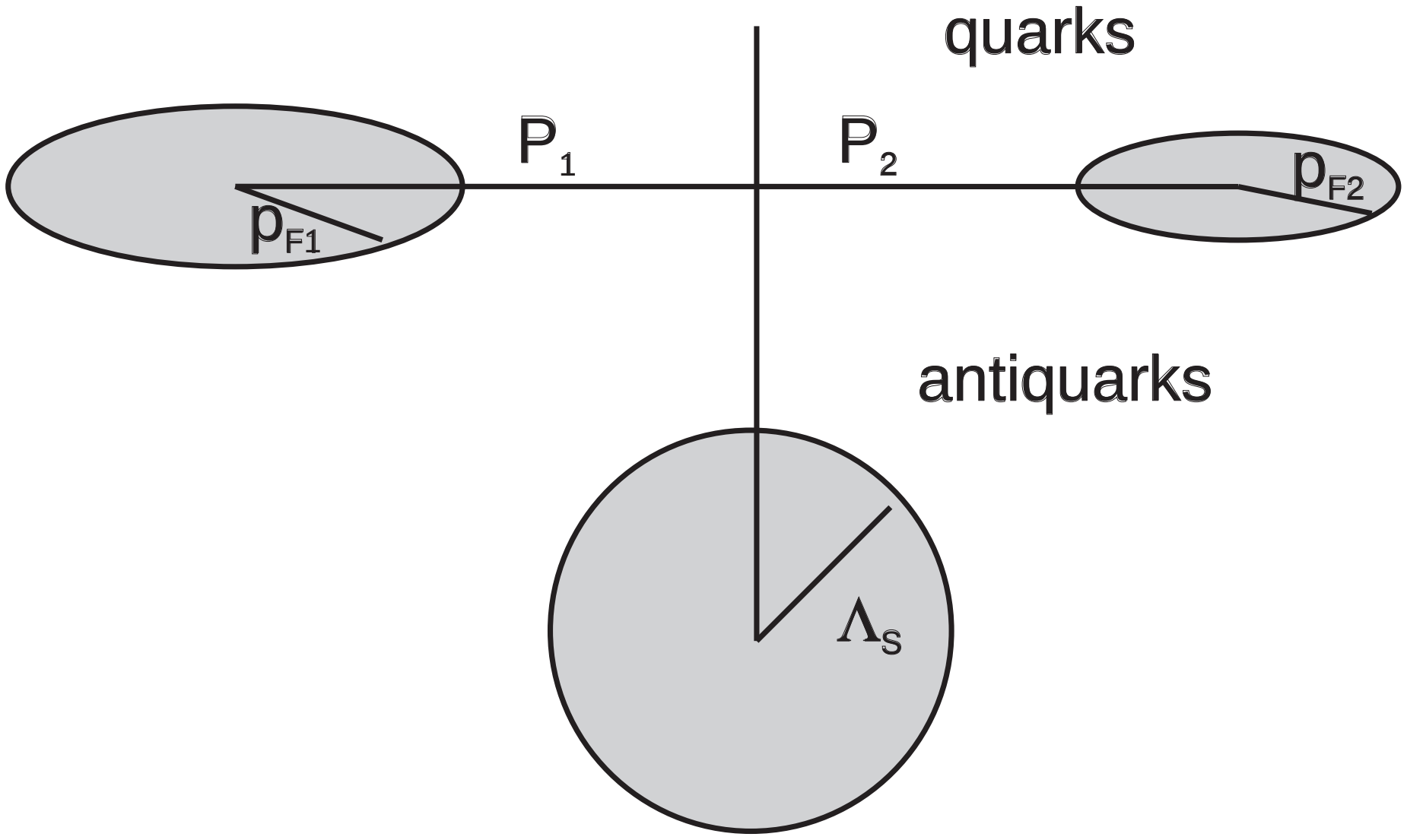

Fig. 2. Characteristic quark phase-space distribution in the overlap regime for two colliding nucleons for $p_y = 0$. The sphere with radius $\Lambda_S$ characterizes the Dirac sea contribution at rest.

Due to the 3-momentum cutoff $\Lambda_S$ in the gap equation (13) our prescription is not Lorentz-invariant and all quantities computed depend on the reference frame. For nucleon-nucleon collisions the natural frame of reference is the c.m.s., i.e. $P_1 = P_2$, which should be small compared to the nucleon mass. For the following illustration we thus restrict to small relative momenta of the 'nucleons' $P = 3(P_1 + P_2) = 6P_1$.

Iteration of the gap equation (13) for the two nucleon system then yields the effective mass $m_u(\mathbf{r}; P_1, P_1, \rho^1, \rho^2)$ as well as the scalar density $\rho_S(\mathbf{r})$. The resulting quark vector (solid lines) and scalar densities (dashed lines) are dis-played in Fig. 3 as a function of $z = r$ for $x = y = 0$ for different distances $R$ of the two nucleons (and a constant relative momentum $P = 0.2$ GeV/c). Due to the gap equation (13) the scalar quark density drops substantially even for a moderate overlap of the nucleons ($R \approx 1.6$fm), which reflects the 'intermediate' range ($\sigma$-field) attraction of the two nucleons, whereas the overlap of the vector densities becomes more substantial at short distance ($R \approx 0.5$fm), which reflects the $\omega$-field in terms of the conventional $\sigma$-$\omega$ model [40].

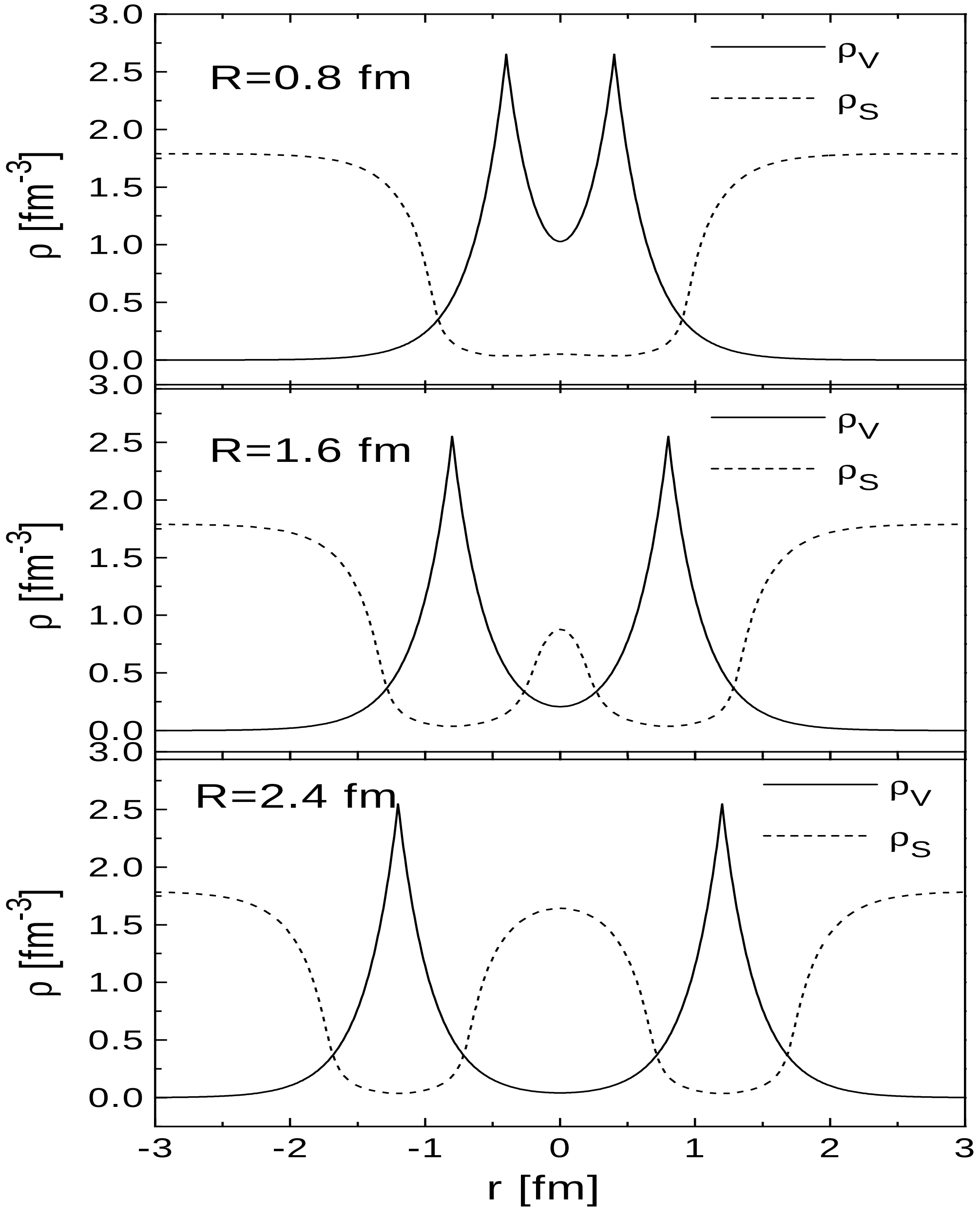

Fig. 3. Spatial quark distribution $2/3\ \rho_u(x = 0, y = 0, z = r)$ (full lines) and scalar quark density (dashed lines) for two colliding nucleons with relative momentum $P = 0.2$ GeV/c for different relative distances $R$ from 0.8 fm to 2.4 fm.

We note that due to the non-covariant cutoff $\Lambda_{\rm S}$ in the gap equation (13) the effective mass $m_{\rm u}$ of a quark becomes momentum dependent for nucleon-nucleon configurations even in the c.m.s. This is more quantitatively shown in Fig. 4 where $m_{\rm u}(\rho_{\rm u}, P)$ is displayed as a function of $P_1 = P_2 = P$ and $\rho^1 = \rho^2 = \rho_{\rm u}$ to illustrate the smooth general dependence on density and relative momentum. It is clearly seen from Fig. 4 that the effective mass drops with density $\rho$ and increases for fixed $\rho$ with the relative momentum $P_{\rm r} = 2P$. Since the origin of this momentum dependence is not of dynamical nature - e.g. a finite range of the scalar quark-quark interaction - one has to worry about its consequences for the nuclear-matter computations we aim at. For $\rho_{\rm B} = 10\rho_0$ we get a nuclear Fermi momentum $p_{\rm F}^{\rm N} \approx 0.57$ GeV/c and thus for the relative momentum parameter $P$ we have $P = p_{\rm F}^{\rm N}/3 \approx 0.19$ GeV/c. A closer look at Fig. 4 then tells us that the effective quark mass is nearly independent on $P$ for $P \leq 0.2$ GeV/c such that the momentum dependence of $m_{\rm u}$ is rather insignificant for our purposes.

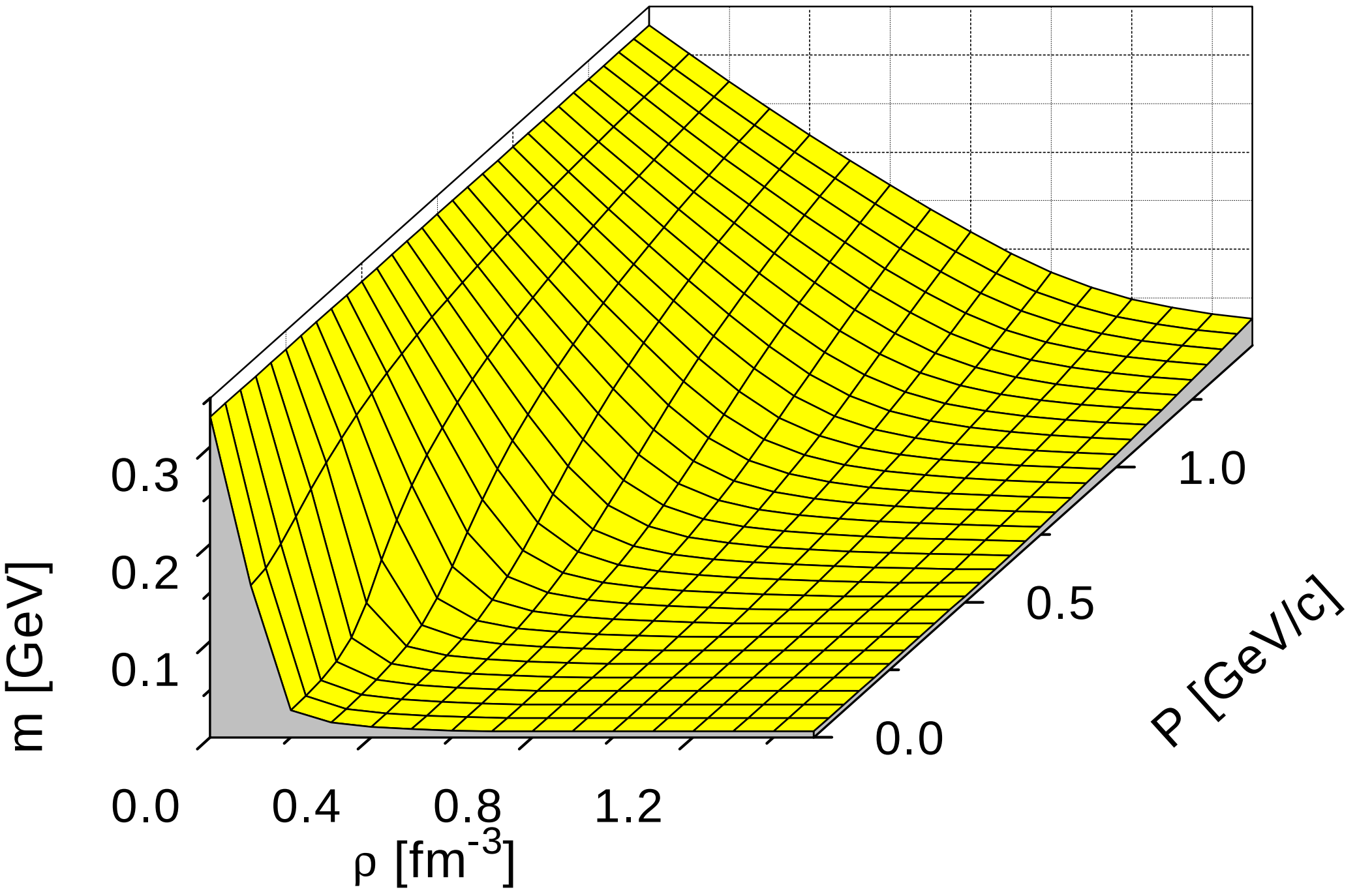

Fig. 4. Effective quark mass $m = m_{\rm u}(\rho_{\rm u}, P)$ as a function of $P = P_1 = P_2$ and $\rho_{\rm u} = \rho^1 = \rho^2$ according to the gap equation (13).

However, before evaluating the energy density for nuclear matter configurations we have to make sure that for vanishing nuclear density the energy of a nucleon moving with momentum $p_{\rm N} = 3p_{\rm u}$ agrees with the dispersion relation of a free nucleon, i.e.

$$E(p_{\rm N}) = \sqrt{p_{\rm N}^2 + M_{\rm N}^2} \ , \tag{24}$$

where $M_{\rm N}$ is the nucleon mass in its rest frame. This is indeed the case as shown in Fig. 5, where the relation (24) (dashed line) is compared to the result from integrating $T^{00}(\mathbf{r})$ over $\mathbf{r}$ (solid line). The slight deviations from the exact result (24) provide a measure for the violation of covariance in the model adopted here which, however, are not serious for nucleon momenta $p_{\rm N} \leq 0.57$ GeV/c. The result in Fig. 5 comes about because the total energy of the nucleon is dominated by the relative kinetic energy of the quarks.

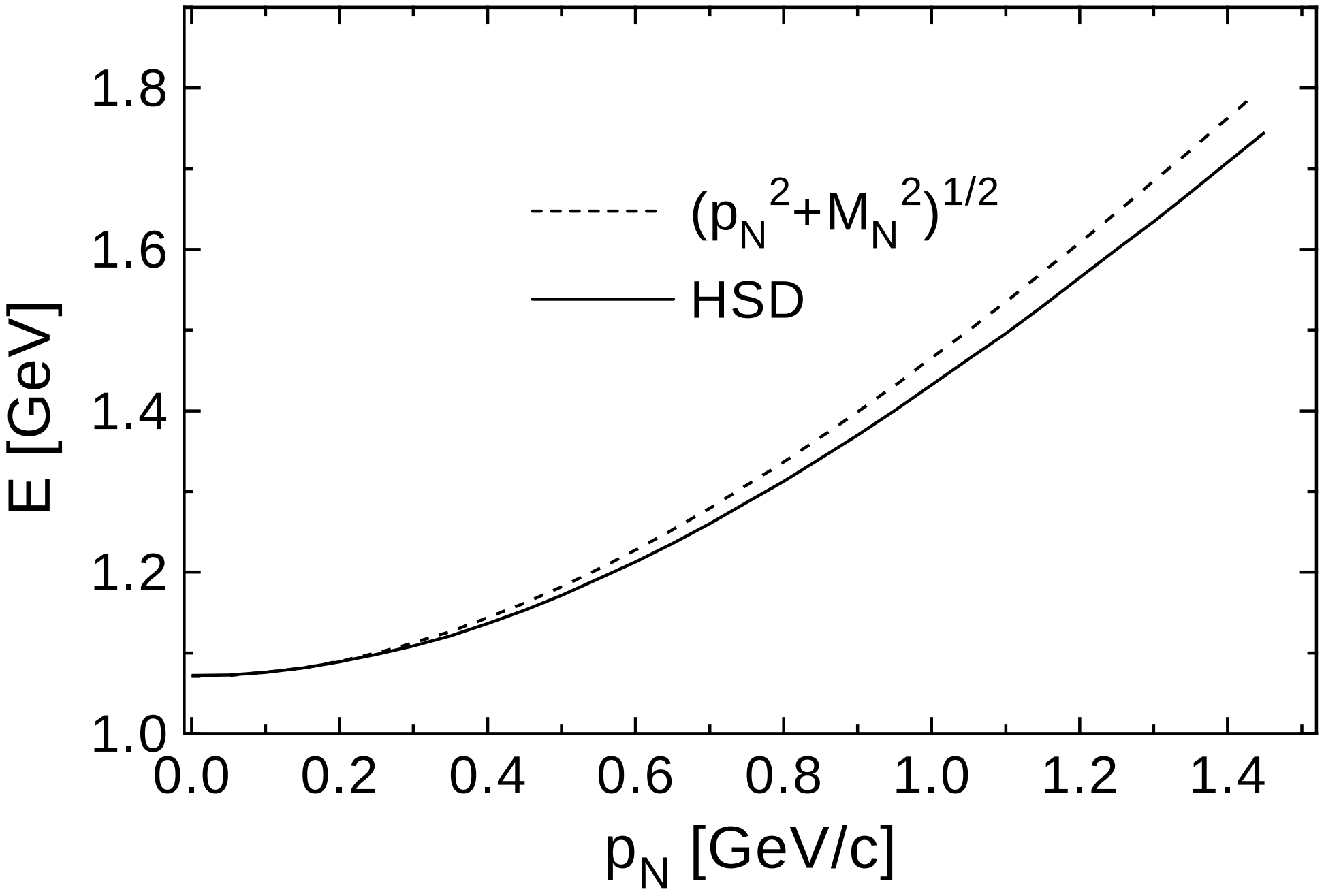

Fig. 5. Free nucleon dispersion relation (24) (dashed line) and $\int d^3r\ T^{00}(\mathbf{r})$ (solid line) for a 'nucleon' moving with momentum $P_{\rm N}$.

We now continue with the isospin symmetric nuclear matter problem where the nucleon phase-space distribution for fixed spin and isospin is given by

$$f_{\rm N}(\mathbf{r}_{\rm N}, \mathbf{p}_{\rm N}) = \Theta(p_{\rm F} - \mid \mathbf{p}_{\rm N} \mid) \ , \tag{25}$$

with the nucleon Fermi momentum $p_{\rm F} = (6/4\pi^2 \rho_{\rm N})^{1/3}$, where $\rho_{\rm N}$ is the nuclear matter density which will be discussed in units of $\rho_0 \approx 0.17$ fm$^{-3}$.

Here a further problem is related with the change of the nucleon formfactor in the medium. As suggested e.g. by the interpretation of the EMC effect by Close et al. [49] or arguments based on chiral symmetry by Brown [16] the nucleon might change its size in the nuclear medium such that the vector density of a quark is no longer given by (16). A fully dynamical model of the nucleon in the nuclear medium should give this modification of the formfactor in a selfconsistent manner. Since the Lagrangian (8) here is only considered to provide an effective quark-quark interaction for the energy density (21) we model such in-medium effects by modifying the width parameter $b_0$ in (16) as

$$b_0(\rho_{\rm N}) = 0.25\,{\rm fm}\left(1 + \alpha\;\frac{\rho_{\rm N}}{\rho_0}\right) \tag{26}$$

with a parameter $\alpha$ to be determined by the nuclear matter saturation point (see below).

In order to carry out computations for the nuclear matter problem we simulate the quark phase-space distribution $f_{\rm u}(\mathbf{r},\mathbf{p})$ - which enters $T^{00}$ in (21) - by characteristic samples $k$

$$f_{\rm u}^k(\mathbf{r},\mathbf{p}) = \sum_{j=1}^{A} f_{\rm u}(\mathbf{r}-\mathbf{r}_{\rm N}^j, \mathbf{p}-\mathbf{p}_{\rm N}^j; b_0(\rho_N)) \;\;, \tag{27}$$

where $f_{\rm u}(\mathbf{r},\mathbf{p},b_0)$ denotes the semiclassical quark phase-space distribution for a 'nucleon' of width $b_0(\rho_{\rm N})$ (26). The nucleon positions $\mathbf{r}_{\rm N}^j$ are determined by Monte Carlo in a box of volume $V = a^3$ with $-a/2 \leq x_{\rm N}^j, y_{\rm N}^j, z_{\rm N}^j \leq a/2$. Only those samples are accepted for which the average distance to the next neighbour agrees within 3% with that for the respective infinite nuclear matter value. The nucleon momenta $\mathbf{p}_{\rm N}^j$ then are selected by Monte Carlo with the constraint $|\mathbf{p}_{\rm N}^j| \leq p_{\rm F}(\rho_{\rm N})$ and $\sum_j \mathbf{p}_{\rm N}^j = 0$. Additionally we rejects samples where the average kinetic energy

$$T_{\rm N} = \frac{1}{A}\sum_{j=1}^{A}\left(\sqrt{(\mathbf{p}_{\rm N}^j)^2 + M_{\rm N}^2} - M_{\rm N}\right) \tag{28}$$

does not match with the nuclear matter value within 3%. The density $\rho_{\rm N}$ in these simulations is given by $\rho_{\rm N} = A/V$ (input) while $A = 64$ has been adopted throughout the calculations. In order to compile the dependence of the total energy on $\rho_{\rm N}$ we have scaled the individual positions $\mathbf{r}_{\rm N}^j$ with $a \sim \rho_{\rm N}^{-1/3}$ and the momenta $\mathbf{p}_{\rm N}^j$ with $\rho_{\rm N}^{1/3}$.

A snapshot of the quark density (for fixed z) for a chacteristic sample $k$ at normal nuclear matter density $\rho_0$ is shown in the upper part of Fig. 6; the resulting effective mass $m_{\rm u}(\mathbf{r})$ according to the gap equation (13) - for the configuration shown in the upper part of Fig. 6 - is displayed in its lower part. Since at normal nuclear matter density the overlap of the nucleons is only moderate, the individual scalar 'quark bags' can still approximately be separated in space for a given time. As an example for higher nucleon density

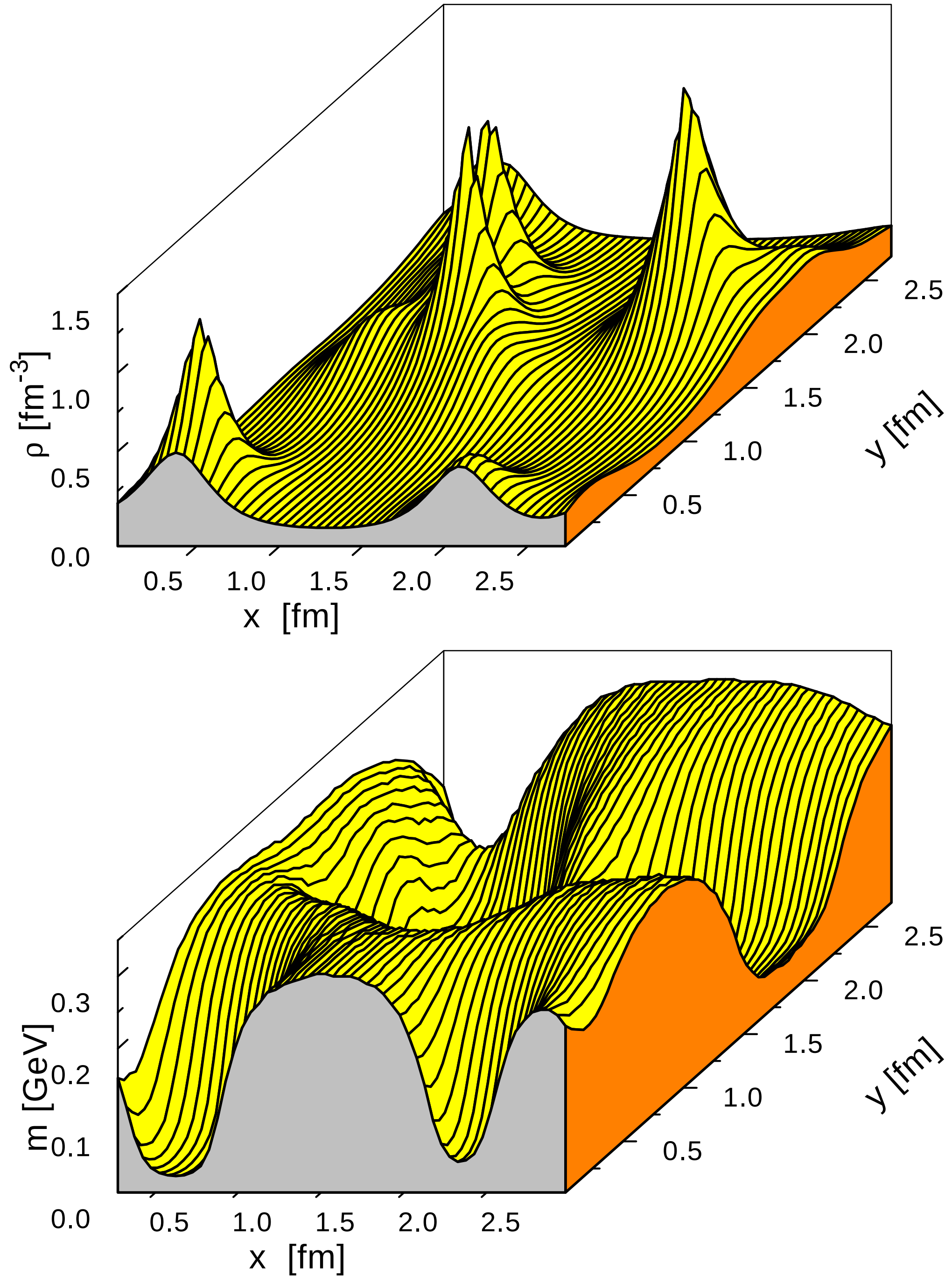

Fig. 6. Snapshot of the spatial quark distribution (for fixed z) at normal nuclear matter density $\rho_0$ (upper part) for $\alpha = 0.18$ together with the resulting effective quark mass $m = m_{\rm u}(x, y)$ (lower part).

we show a snapshot of the quark distribution at $4 \times \rho_0$ for $\alpha = 0.18$ in Fig. 7

(upper part) together with the corresponding quark mass $m_{\rm u}(x, y, z = \text{const})$ (lower part) from the gap equation (13). Since the overlap of the quark distributions now becomes substantial, the average quark mass drops to about 30 MeV indicating partial chiral symmetry restoration.

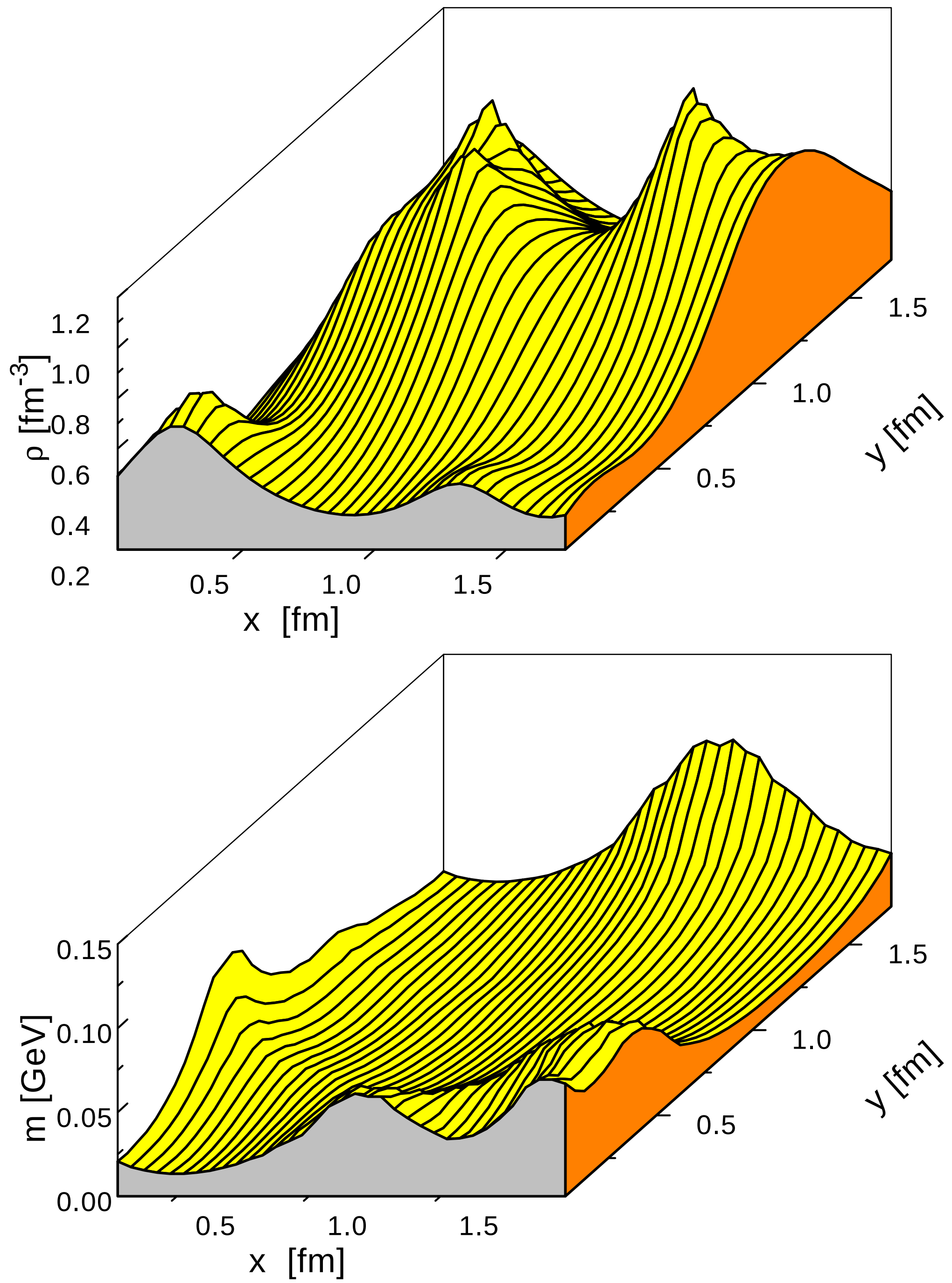

Fig. 7. Snapshot of the spatial quark distribution (for fixed z) at $4 \times \rho_0$ (upper part) together with the resulting effective quark mass $m = m_{\rm u}(x, y)$ (lower part).

Now performing the integration of $T^{00}(\mathbf{r})$ over coordinate space and averaging over characteristic samples $k$ for nuclear matter configurations (as shown in Fig. 6) [5], dividing by the number of nucleons on the grid and subtracting the bare nucleon mass we can compute the energy per nucleon ($N_\mathrm{k} \approx 100$),

$$\frac{E}{A}\left(\frac{\rho}{\rho_0}\right) = \frac{1}{A}\frac{1}{N_\mathrm{k}}\sum_{k=1}^{N_\mathrm{k}} \int d^3r \; T_k^{00}(\mathbf{r}) - M_\mathrm{N}, \tag{29}$$

and thus establish a direct link between the energy density of quarks with the energy per nucleon of isospin symmetric nuclear matter at finite density $\rho/\rho_0$. In (29) the energy density $T_k^{00}(\mathbf{r})$ is defined by eq. (21) with $f_\mathrm{u}(\mathbf{r},\mathbf{p})$ replaced by $f_\mathrm{u}^k(\mathbf{r},\mathbf{p})$ from (27).

The energy per nucleon (29) (for $\alpha = 0.18$) is shown in Fig. 8 (full line) in comparison to the Dirac-Brueckner results from [13] (full squares) and the parametrizations POL6 and POL7 of the RBUU approach [9] that were found to optimally describe heavy-ion reactions in the energy regime up to about 1 GeV/A. We find the binding energy per nucleon ($\approx -16$ MeV at $\rho_\mathrm{N} = \rho_0$) to be reproduced well for $\alpha \approx 0.18$ which corresponds to a swelling of the nucleon by 18% at normal nuclear matter density. For $\alpha = 0$ there is no minimum in E/A due to the Pauli pressure such that the swelling of the nucleon - which enhances the scalar attraction and reduces the vector repulsion - is a necessary phenomenon within the present approach to achieve proper binding. The resulting incompressibility $K$ of nuclear matter amounts to about $K \approx 250$ MeV.

Since the energy per nucleon in our approach (HSD) is well in between the limits of POL6 and POL7, as extracted from detailed comparisons in ref. [9] for nucleus-nucleus collisions in the SIS energy regime, we infer that the equation of state generated by the model is quite realistic in the lower density ($\rho \leq 3\rho_0$) regime. Its extension to $10\rho_0$ (lower part of Fig. 8), however, might still be questionable and has to be examined in comparison to experimental flow data at much higher (e.g. AGS) bombarding energies.

The nuclear equation of state (EOS) in Fig. 8 shows no density isomer up to $10\rho_0$ on the basis of the effective quark model adopted. The thermodynamic

[5] For technical reasons we first look for the 'nucleon' that exhibits a maximum quark density at a given grid point $\mathbf{r}$ (giving $\rho^1, P_1$) and then sum up the quark contributions of the other 'nucleons' (giving $\rho^2, P_2$). The corresponding values for $m_\mathrm{u}(\mathbf{r})$ and $T^{00}(\mathbf{r})$ are then taken from the parametrized configurations displayed in Fig. 2. We have tested for a couple of samples that this approximate evaluation works quite well if ensemble averages for nuclear matter configurations are considered.

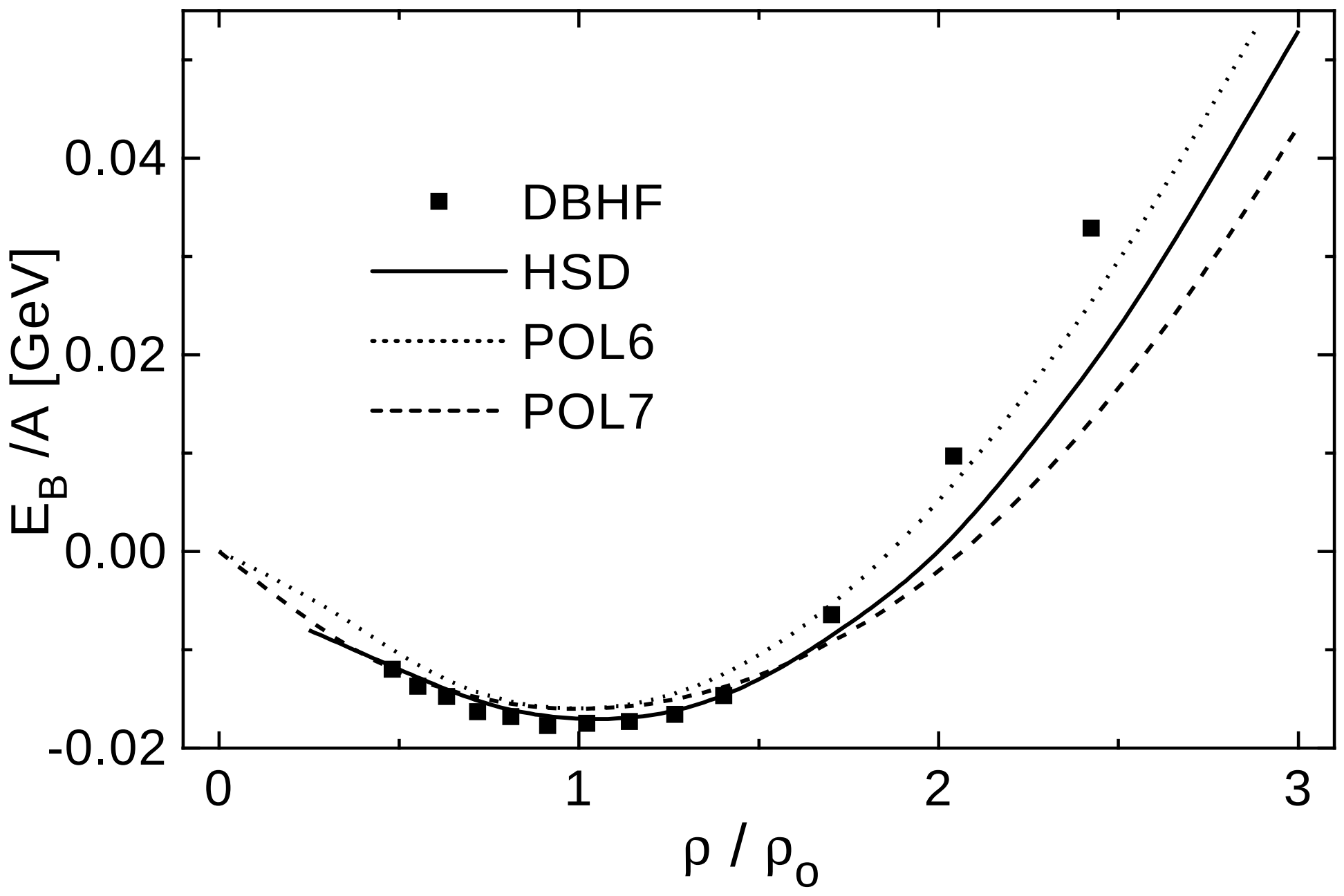

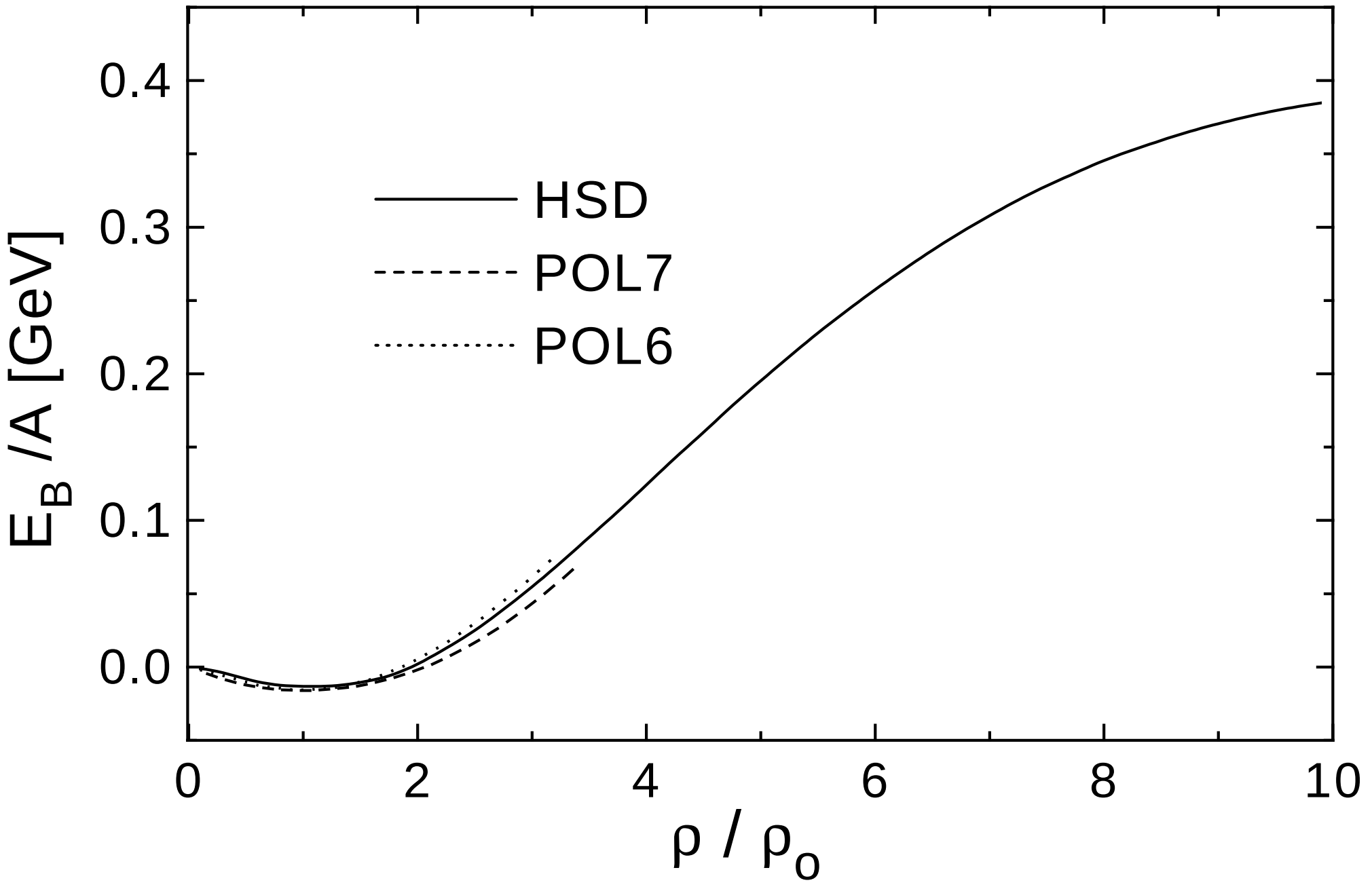

Fig. 8. Equation of state for nuclear matter; HSD (solid line), DBHF (full squares); RBUU results: POL6 (dotted line), POL7 (dashed line) from ref. [9].

pressure

$$P_{\mathrm{T}} = \rho^2 \frac{\partial}{\partial \rho} \left( \frac{E}{A} \right) \quad , \tag{30}$$

furthermore, increases quadratically for $\rho/\rho_0 > 2$ and slightly levels off at high density, but does not drop to zero in the range considered here.

In view of the rather simple shape of the EOS in Fig. 8 and its similarity to the RBUU parameter sets POL6 and POL7 from ref. [9], it is now almost straight forward to 'extract' nucleon selfenergies $U_{\rm N}^{\rm S}$ and $U_{\rm N}^{\mu}$ for the hadronic transport approach (1).

### *2.4 Nucleon selfenergies*

The scalar and vector mean fields $U_{\rm h}^{\rm S}$ and $U_{\rm h}^{\mu}$ in eq. (1) for nucleons now can be specified along the line of ref. [7]. In order to achieve a covariant transport approach, which is also thermodynamically consistent [7], we parametrize the scalar and vector selfenergies in phase-space representation as

$$\begin{aligned} U^{\rm S}(x,p) &= U^{\rm S}_{\rm loc}(x) - \frac{4}{(2\pi)^3}\frac{\bar g_{\rm S}^2}{m_{\rm S}^2}\int d^4p'\, M^*(x,p')\frac{\bar\Lambda_{\rm S}^2}{\bar\Lambda_{\rm S}^2-(p-p')^2}\, f_{\rm N}(x,p')\ , \\ U^{\mu}(x,p) &= U^{\mu}_{\rm loc}(x) + \frac{4}{(2\pi)^3}\frac{\bar g_{\rm V}^2}{m_{\rm V}^2}\int d^4p'\, \Pi^{\mu}(x,p')\frac{\bar\Lambda_{\rm V}^2}{\bar\Lambda_{\rm V}^2-(p-p')^2}\, f_{\rm N}(x,p')\ , \end{aligned} \tag{31}$$

where $f_{\rm N}(x,p)$ is the nucleon phase-space distribution. In (31) the effective nucleon mass $M^*(x,p)$ and the kinetic momentum $\Pi^{\mu}(x,p)$ are given by

$$\begin{aligned} M^*(x,p) &= M_{\rm N} + U^{\rm S}(x,p)\ , \\ \Pi^{\mu}(x,p) &= p^{\mu} - U^{\mu}(x,p)\ . \end{aligned} \tag{32}$$

In (31) $U^{\rm S}_{\rm loc}(x)$ and $U^{\mu}_{\rm loc}(x)$ are the local parts of the selfenergy,

$$\begin{aligned} U^{\rm S}_{\rm loc}(x) &= -g_{\rm S}^0\, \sigma_{\rm H}(x), \\ U^{\mu}_{\rm loc}(x) &= g_{\rm V}^0\, \omega^{\mu}_{\rm H}(x) \end{aligned} \tag{33}$$

with

$$\begin{aligned} \omega^{\mu}_{\rm H}(x) &= \frac{g_{\rm V}^0}{m_{\rm V}^2}\frac{4}{(2\pi)^3}\int d^4p\, \tilde\Pi^{\mu}(x,p) f_{\rm N}(x,p)\ , \\ \tilde\Pi^{\mu} &= \Pi^{\mu} - \Pi^{\nu}(\partial_p^{\mu} U_{\nu}) - M^*(\partial_p^{\mu} U^{\rm S})\ , \end{aligned} \tag{34}$$

while $\sigma_{\rm H}(x)$ is obtained from the solution of

$$m_{\rm S}^2\sigma_{\rm H} + B\sigma_{\rm H}^2 + C\sigma_{\rm H}^3 = g_{\rm S}^0\frac{4}{(2\pi)^3}\int d^4p\, M^*(x,p) f_{\rm N}(x,p). \tag{35}$$

The quasi-particle properties are defined via the mass shell constraint [7]

$$\delta(\Pi_\mu \Pi^\mu - M^{*2}) \ . \tag{36}$$

The associated energy-momentum tensor reads

$$\begin{aligned}
T_{\rm N}^{\mu\nu}(x) &= \frac{4}{(2\pi)^3} \int d^4p \, \tilde{\Pi}^\mu p^\nu f_{\rm N}(x,p) \\
&+ \ (\partial^\mu \sigma_{\rm H}(x))(\partial^\nu \sigma_{\rm H}(x)) - (\partial^\mu \omega_{\rm H}^\lambda(x))(\partial^\nu \omega_\lambda^{\rm H}(x)) \\
&+ \ \left\{ \frac{1}{2} m_{\rm S}^2 \sigma_{\rm H}^2 + \frac{1}{3} B \sigma_{\rm H}^3 + \frac{1}{4} C \sigma_{\rm H}^4 - \frac{1}{2}(\partial_\lambda \sigma_{\rm H})(\partial^\lambda \sigma_{\rm H}) - \frac{1}{2} m_{\rm V}^2 \omega_\lambda^{\rm H} \omega_{\rm H}^\lambda \right. \\
&+ \ \frac{1}{2}(\partial_\lambda \omega_{\rm H}^\delta)(\partial^\lambda \omega_\delta^{\rm H}) \\
&- \ \frac{2}{(2\pi)^3} \int d^4p \, M^*(x,p) \frac{4}{(2\pi)^3} \frac{\bar{g}_{\rm S}^2}{m_{\rm S}^2} \int d^4p' \, M^*(x,p') \frac{\bar{\Lambda}_{\rm S}^2}{\bar{\Lambda}_{\rm S}^2 - (p-p')^2} f_{\rm N}(x,p') \\
&- \ \left. \frac{2}{(2\pi)^3} \int d^4p \, \Pi^\lambda(x,p) \frac{4}{(2\pi)^3} \frac{\bar{g}_{\rm V}^2}{m_{\rm V}^2} \int d^4p' \, \Pi_\lambda(x,p') \frac{\bar{\Lambda}_{\rm V}^2}{\bar{\Lambda}_{\rm V}^2 - (p-p')^2} f_{\rm N}(x,p') \right\} g^{\mu\nu} \ .
\end{aligned} \tag{37}$$

In this hadronic approach with momentum-dependent fields the 'free' parameters $g_{\rm S}^0, \bar{g}_{\rm S}, g_{\rm V}^0, \bar{g}_{\rm V}, m_{\rm S}, m_{\rm V}, \bar{\Lambda}_{\rm S}, \bar{\Lambda}_{\rm V}, B, C$ allow to describe almost arbitrary equations of state and nucleon selfenergies. For nuclear matter at density $\rho_{\rm N}$ the energy per nucleon is given by

$$\frac{E}{A} = \frac{T_{\rm N}^{00}}{\rho_{\rm N}} - M_{\rm N} \ , \tag{38}$$

where $M_{\rm N}$ denotes the bare nucleon mass. The evaluation of $T_{\rm N}^{00}$ for

$$f_{\rm N}(p) = 2\Theta(\Pi_0)\delta(\Pi^2 - M^{*2})\Theta(p_{\rm F}^{\rm N} - |\mathbf{p}|) \tag{39}$$

with the nucleon Fermi momentum $p_{\rm F}^{\rm N}$ then reduces to the coupled eqs. (44)-(49) in ref. [7] which don't have to be repeated here.

The key link for determining the free parameters in the hadronic model above now is the *model independent* relation for the effective quark mass as a function of (small) $\rho_{\rm N}$

$$m_{\rm u}(\rho_{\rm N}) = m_V \left( 1 - \frac{\Sigma_{\pi N}}{f_\pi^2 m_\pi^2} \rho_{\rm N} \right) \ , \tag{40}$$

which follows from the Hellmann-Feynman theorem and the GOR relation (12) [50]. In (40) $m_{\rm V}$ is the vacuum effective quark mass from (11). In this context we show in Fig. 9 the average effective quark mass (in units of the vacuum mass $m_V$) (solid line) as a function of the nuclear density $\rho_N = \rho$ as obtained from the nuclear matter simulations. The effective quark mass drops by about 35 % at $\rho_0$ according to (40) with $\sum_{\pi \mathrm{N}} = 47$ MeV, and essentially continues with a constant slope up to about $2 \times \rho_0 \approx 0.33 \mathrm{fm}^{-3}$ in line with the Dirac-Brueckner analysis in ref. [51](cf. also ref. [52]). The bare quark mass then is reached at about $\rho_{\mathrm{N}} \approx 0.6 \ \mathrm{fm}^{-3}$. It is important to note that the effective nucleon mass $M^*(p_{\mathrm{F}}^{\mathrm{N}})$ (normalized to the vacuum mass) in the RBUU approach of ref. [9] shows the same scaling with density up to about $\rho_0$, which is also well in line with Dirac phenomenology.

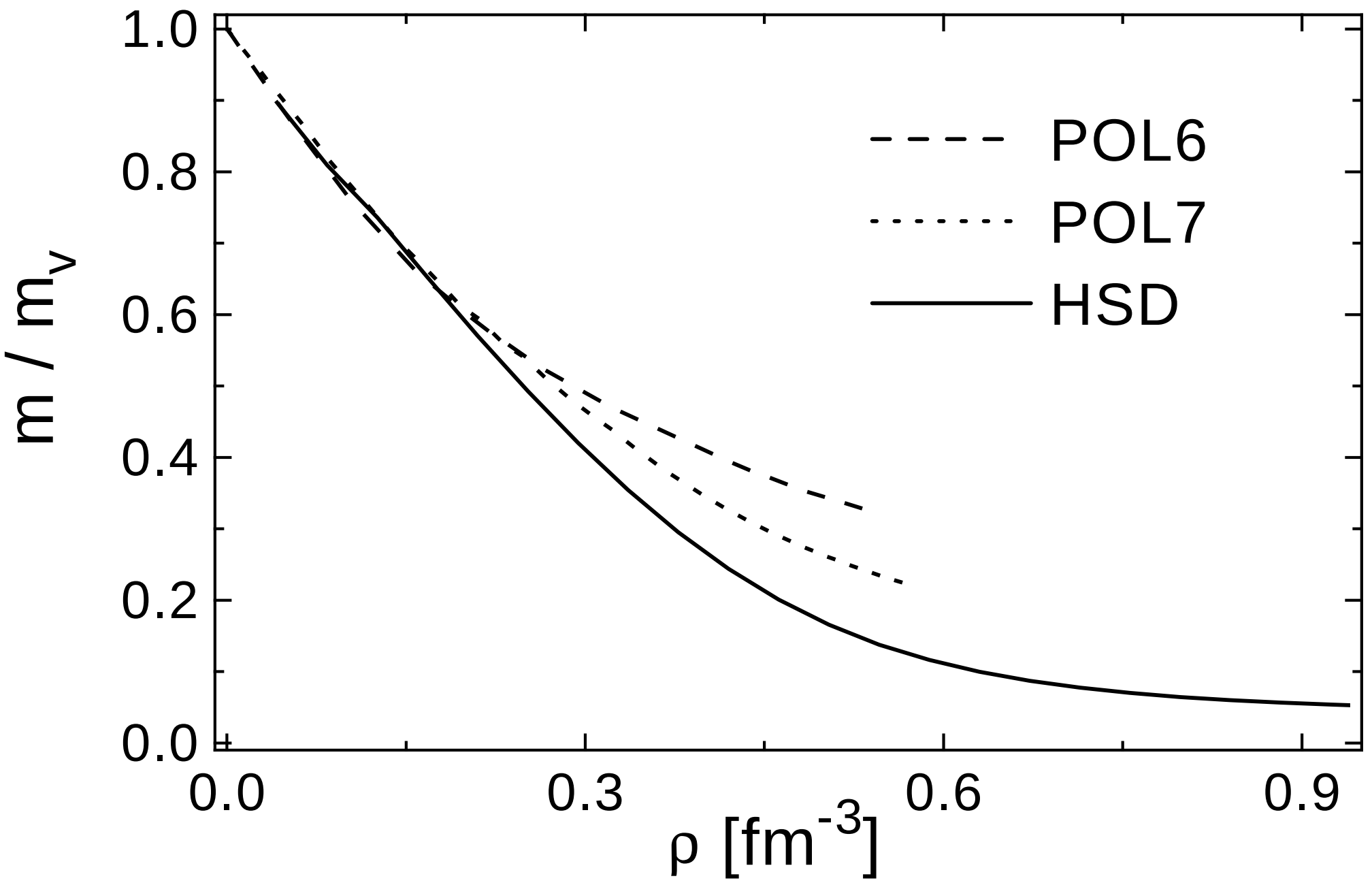

Fig. 9. Effective mass divided by the vacuum mass as a function of the nucleon density $\rho_{\mathrm{N}} = \rho$; quark mass $m = m_{\mathrm{u}}(\rho_N)$ in the HSD approach (solid line); nucleon mass in the RBUU approach: POL6 (dotted line), POL7 (dashed line).

Thus observing that the equation of state from the effective NJL model (Fig. 8) as well as the relative scaling of the quark mass with nucleon density $\rho_{\mathrm{N}}$ (Fig. 9) is very similar to the more traditional RBUU transport approach from refs. [7,9] at low density $\rho_{\mathrm{N}}$, we fix the parameters $g_{\mathrm{S}}^0, \bar{g}_{\mathrm{S}}, g_{\mathrm{V}}^0, \bar{g}_{\mathrm{V}}, \bar{\Lambda}_{\mathrm{S}}, \bar{\Lambda}_{\mathrm{V}} \ldots$ by the condition

$$\frac{M^*(\rho_{\mathrm{N}}, p_{\mathrm{F}}^{\mathrm{N}})}{M_{\mathrm{N}}} = \frac{m_{\mathrm{u}}(\rho_{\mathrm{N}})}{m_{\mathrm{V}}} \quad , \tag{41}$$

which essentially determines the scalar selfenergy of the nucleon, as well as the equation of state from Fig. 8,

$$\left(\frac{E}{A}\right)_{\rm HSD} = \frac{T^{00}_{\rm N}}{\rho_{\rm N}} - M_{\rm N} \tag{42}$$

up to $\rho_{\rm N} = 10\rho_0$. The scalar and vector nucleon selfenergies then are uniquely determined by eqs. (31) - (39) for arbitrary nucleon phase-space distributions.

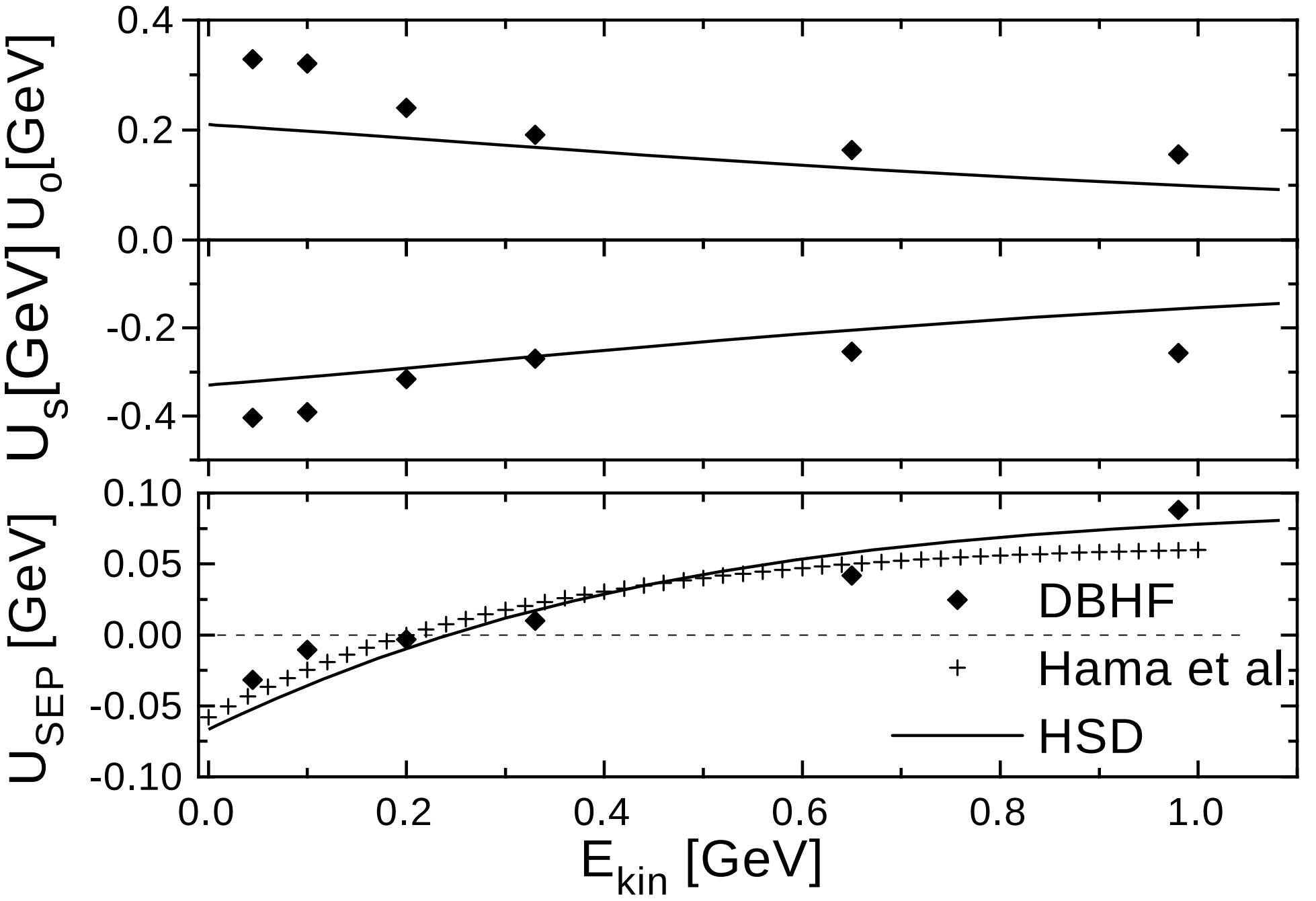

Fig. 10. Nucleon selfenergies $U_{\rm S}$, $U_0$ and the Schroedinger equivalent potential $U_{\rm SEP}$ as a function of the nucleon kinetic energy $E_{\rm kin}$ with respect to the nuclear matter rest frame. HSD (solid line); DBHF (full squares); exp. data from Hama et al. [53] (crosses).

In Fig. 10 we compare the resulting momentum dependence of the nucleon selfenergies at density $\rho_0$ with Dirac-Brueckner results from [13] (full dots). In the lower part of Fig. 10 the real part of the Schroedinger equivalent potential (SEP)

$$U_{\rm SEP} = U_{\rm S}(\rho_0, P) + U_0(\rho_0, P) + \frac{1}{2M_{\rm N}}(U_{\rm S}(\rho_0, P)^2 - U_0(\rho_0, P)^2)$$
$$+ U_0(\rho_0, P)\frac{\sqrt{P^2 + M_{\rm N}^2} - M_{\rm N}}{M_{\rm N}} \tag{43}$$

is additionally shown (full line) in comparison to the optical potential analysis from Hama et al. [53] (dashed line) and Dirac-Brueckner computations from [13] up to momenta of 1 GeV/c. This comparison shows that the overall properties of the nucleon selfenergies for $\rho_N \leq 3\rho_0$ and $E_{kin} \leq 1$ GeV are reasonably met by our approach.

Apart from the close analogy of our results with the $\sigma$-$\omega$ model at 'low' momenta (cf. Fig. 10) we are especially interested in the 'high' momentum properties of the present approach, where the standard $\sigma$-$\omega$ model is known to fail significantly. The respective results from our present approach for the scalar and vector nucleon selfenergy as well as the Schroedinger equivalent optical potential in analogy to Fig. 10 are displayed in Fig. 11 up to relative kinetic energies of 15 GeV. Whereas the scalar and vector nucleon selfenergies are found to gradually decrease with momentum (or kinetic energy) - which is essentially a consequence of the cutoff $\Lambda_V \approx 1.5$ GeV introduced in eq. (20) - the Schroedinger equivalent potential exhibits a maximum of about 70 MeV at 1 GeV and drops again for higher kinetic energy. Thus we expect the effects from the real part of the nucleon selfenergies to be of minor importance in the initial phase of nucleus-nucleus collisions at bombarding energies of a few GeV/A, where nucleon cascading with inelastic nucleon excitations should be dominant, i.e. the imaginary part of the hadron selfenergies (cf. Section 3).

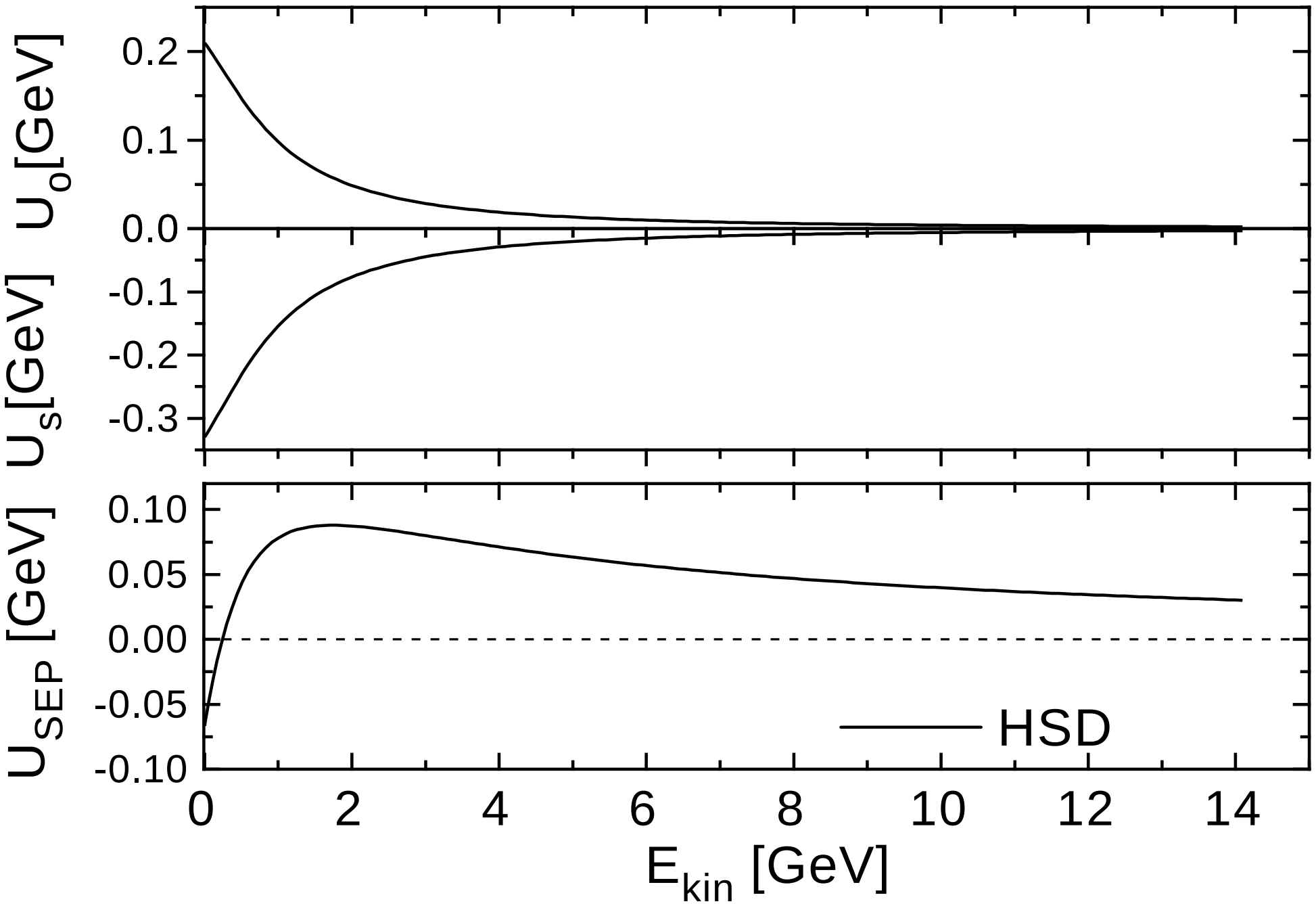

Fig. 11. Nucleon selfenergies $U_S$, $U_0$ and the Schroedinger equivalent potential $U_{SEP}$ as a function of the nucleon kinetic energy $E_{kin}$ at normal nuclear matter density $\rho_0$.

Since a transport approach for high energy nucleus-nucleus collisions also has to include excited states of the nucleon as well as hyperons - we include nucleons, $\Delta$'s, N*(1440), N*(1535), $\Lambda$ and $\Sigma$ hyperons as well as their antiparticles - their respective selfenergies have to be specified, too. As a first approximation we assume here that all baryons (made out of light (u,d) quarks) have the same scalar selfenergies as the nucleons; the vector selfenergy for antiparticles is introduced with a relative (-) sign according to time reversal[6] while the hyperons pick up a factor 2/3 according to the light quark content.

### 2.5 *Meson selfenergies*

Whereas the baryon selfenergies $U_{\rm h}^{\rm S}$ and $U_{\rm h}^{\mu}$ are a necessary ingredient for a relativistic transport model to achieve a realistic description of finite nuclei and intermediate energy nucleus-nucleus reactions, the meson selfenergies might be neglected in zero'th order as in conventional cascade simulations. However, in order to explore dynamical effects from a phase, where the chiral symmetry might be restored, they have to be specified as well (on the one-loop level) e.g. by a suitable Lagrangian density.

In the HSD approach, where we propagate explicitly pions, kaons, $\eta$'s and the vector mesons $\omega$, $\rho$, $\phi$, and $K^*(892)$ we assume that the pions as Goldstone bosons do not change their properties in the medium; we also discard selfenergies for the $\eta$-mesons in the present version. Thus a Lagrangian density for the coupled system of baryons and mesons can be written as

$$\mathcal{L}_{\rm H} = \mathcal{L}_{\rm B} + \sum_m \mathcal{L}_m^0 + \mathcal{L}_{\rho {\rm B}}^{\rm int} + \mathcal{L}_{\omega {\rm B}}^{\rm int} + \mathcal{L}_{\phi {\rm B}}^{\rm int} + \mathcal{L}_{\rm KB}^{\rm int} \quad , \tag{44}$$

where $\mathcal{L}_{\rm B}$ corresponds to the baryon Lagrangian (density) specified in Subsection 2.4, $\mathcal{L}_m^0$ is the free meson Lagrangian density for a meson of type $m$ and $\mathcal{L}_{m{\rm B}}^{\rm int}$ denote the meson-baryon interaction densities. The problem now is to fix $\mathcal{L}_{m{\rm B}}^{\rm int}$ in connection with chiral symmetry restoration.

Kaplan and Nelson [24] have shown a way how to proceed in this case. Starting from a $SU(3)_{\rm L} \times SU(3)_{\rm R}$ chiral Lagrangian and using chiral perturbation theory they write down an effective meson-baryon Lagrangian which they claim to be valid up to $\sim 7\rho_0$. Since the coefficients in this Lagrangian are approximately known experimentally (within an uncertainty of about 30%), one can model a Lagrangian of lower complexity, but with the same properties on the mean-field level. Such limits lead to the following dispersion relation for kaons in the nuclear medium [54]:

[6] This limit has to be taken with care because Teis et al. found in [3] that a sign change of the vector potential results in a too strong attraction for antiprotons.

$$\omega_{K^+}(\mathbf{p}) = \left\{ \mathbf{p}^2 + m_K^2 \left( 1 - \frac{\Sigma_{KN}}{f_K^2 m_K^2} \rho_S + \left( \frac{3\rho_B}{8 f_K^2 m_K} \right)^2 \right) \right\}^{1/2} + \frac{3}{8} \frac{\rho_B}{f_K^2} \ ,$$

$$\omega_{K^-}(\mathbf{p}) = \left\{ \mathbf{p}^2 + m_K^2 \left( 1 - \frac{\Sigma_{KN}}{f_K^2 m_K^2} \rho_S + \left( \frac{3\rho_B}{8 f_K^2 m_K} \right)^2 \right) \right\}^{1/2} - \frac{3}{8} \frac{\rho_B}{f_K^2} \ , \qquad (45)$$

with $m_K$ denoting the bare kaon mass, $f_K \approx 93$ MeV and $\Sigma_{KN} \approx 350$ MeV, while $\rho_S$ and $\rho_B$ are the scalar and vector baryon densities, respectively. We thus approximate the interaction density for the $K^+$-baryon system by

$$\mathcal{L}^{int}_{K^+B} = \left\{ \frac{\Sigma_{KN}}{f_K^2} \bar{B}B - \left( \frac{3}{8 f_K^2} \right)^2 \left( \bar{B}\gamma^\mu B \right)^2 \right\} K^{+\dagger} K^+$$
$$+ i \frac{3}{8 f_K^2} (\bar{B}\gamma_\mu B) K^{+\dagger} (\partial^\mu K^+) \ + i \frac{3}{8 f_K^2} (\bar{B}\gamma^\mu B) (\partial_\mu K^{+\dagger}) K^+ \ , \qquad (46)$$

where $\bar{B}, B$ denote the baryon spinors, and for the $K^-$-baryon system by

$$\mathcal{L}^{int}_{K^-B} = \left\{ \frac{\Sigma_{KN}}{f_K^2} \bar{B}B - \left( \frac{3}{8 f_K^2} \right)^2 \left( \bar{B}\gamma^\mu B \right)^2 \right\} K^{-\dagger} K^-$$
$$- i \frac{3}{8 f_K^2} (\bar{B}\gamma_\mu B) K^{-\dagger} (\partial^\mu K^-) \ - i \frac{3}{8 f_K^2} (\bar{B}\gamma^\mu B) (\partial_\mu K^{-\dagger}) K^- \qquad (47)$$

as well as

$$\mathcal{L}^{int}_{K^0B} = \frac{\Sigma_{KN}}{f_K^2} (\bar{B}B) K^{0\dagger} K^0 \qquad (48)$$

with the kaon fields $K^+$, $K^-$ and $K^0$. We assume the same form of $\mathcal{L}^{int}$ for the $K^*$-mesons, too. We note that Li and Ko have recently performed studies on the kaon dynamics at SIS energies with the same type of kaon-baryon interaction [55].

With respect to the interaction of the vector mesons $\rho, \omega, \phi$ with baryons we model $\mathcal{L}^{int}_{mB}$ according to the QCD sum rule studies by Hatsuda and Lee [56] as

$$\mathcal{L}^{int}_{\rho B} = \left( \frac{\lambda_\rho}{\rho_0} \right)^2 m_\rho^2 \left( \bar{B}\gamma^\mu B \right)^2 \rho^{\nu\dagger} \rho_\nu \ ,$$

$$\mathcal{L}^{int}_{\omega B} = \left( \frac{\lambda_\omega}{\rho_0} \right)^2 m_\omega^2 \left( \bar{B}\gamma^\mu B \right)^2 \omega^{\nu\dagger} \omega_\nu \ ,$$

$$\mathcal{L}^{\rm int}_{\phi {\rm B}} = \left(\frac{\lambda_\phi}{\rho_0}\right)^2 m_\phi^2 \left(\bar{\rm B}\gamma^\mu {\rm B}\right)^2 \phi^{\nu\dagger}\phi_\nu \quad , \tag{49}$$

with $\lambda_\rho = \lambda_\omega \approx 0.18$ and $\lambda_\phi \approx 0.025$ in order to obtain a linear dependence of the effective meson masses on the baryon density $\rho_{\rm B} = \langle \bar{\rm B}\gamma^0 {\rm B}\rangle$.

With the specification of the meson-baryon interaction densities $\mathcal{L}^{\rm int}_{m{\rm B}}$ the real part of the hadron selfenergies is now fully defined on the one-loop level. While the determination of the baryon selfenergies was quite involved in this Section in order to achieve valid approximations at low and intermediate energies as well as reasonable extrapolations to high baryon densities, the meson sector still is rather poor and will have to be improved in future.

## 3 Elastic and inelastic hadron scattering

Whereas in a fully selfconsistent relativistic transport theory the real part and the imaginary part of hadron selfenergies are related by means of dispersion relations [1,3,13], it is not justified to employ the model selfenergies (determined in Section 2) in dispersion integrals for the imaginary part because the inelastic scattering rate of nucleons and mesons turns out to be wrong in the limit of vanishing baryon density. As known from transport studies at energies below 2 GeV/A the elementary cross sections in eq. (1) may be approximated by their values in free space. Thus as a first step we adopt the same strategy and use the explicite cross-sections as in the BUU model [57] (for $\sqrt{s} \leq 2.6$ GeV) - that have been successfully tested in the energy regime below 2 GeV/A bombarding energy - and by the LUND string formation and fragmentation model [22] (for $\sqrt{s} > 2.6$ GeV) in case of baryon-baryon collisions. For meson-baryon reactions we adopt a transition energy of $\sqrt{s} = 1.8$ GeV between the known low energy cross sections and the LUND model. We note that the actual values for the transition energies in the elementary cross sections are not sensitive to nucleus-nucleus collisions in the energy regime to be discussed in Section 4.

In order to obtain a rough idea about the inelastic cross sections from the LUND string fragmentation model, we show in Fig. 12 the rapidity spectra for baryons, pions, kaons, $\rho$ and $\omega$ mesons from pp collisions at $T_{\rm lab} = 20$ GeV in the pp center-of-mass system. Whereas the baryons turn out to be located in rapidity close to the initial rapidity of the two colliding baryons, the meson rapidities are dominantly centered around midrapidity with a small contribution from the deexcitation of the baryonic constituents close to the incoming baryon rapidities. This general tendency has to be kept in mind when comparing to nucleus-nucleus collisions later on.

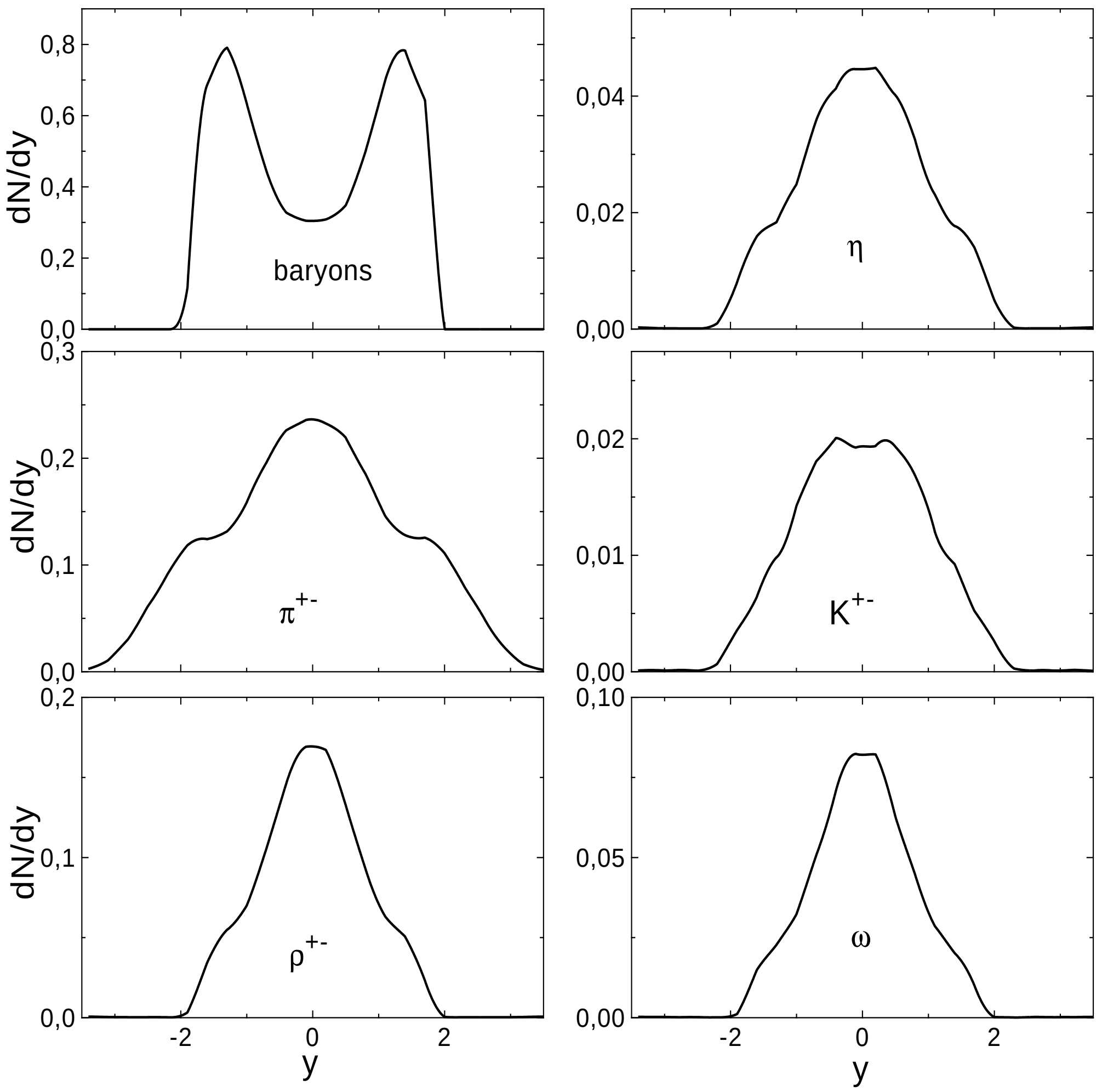

Fig. 12. Rapidity distributions for baryons, pions, kaons, $\eta, \omega$ and $\rho$ mesons from the LUND string fragmentation model for pp collisions at $T_{\rm lab}$ = 20 GeV.

The implementation of the LUND string formation and fragmentation model [22] - which describes the free transition probabilities - in a covariant transport theory implies to use a time scale to transform the cross-sections to collision rates and particle production rates. An appropriate time scale is given by a string formation time $T_{\rm F}$, which denotes the time between the formation and fragmentation of the string in the individual hadron-hadron center-of-mass system for a particle of rapidity $y_{\rm cm} = 0$. Due to covariance this time should be also related to the spatial extension of the interacting hadrons which on average gives $T_{\rm F} \approx 0.8$ fm/c. The sensitivity of the proton rapidity spectra $dN/dy$ to the actual value of $T_{\rm F}$ ist shown in Fig. 13 for central collisions of $^{40}$Ca + $^{40}$Ca at 30 GeV/A. It is seen that $T_{\rm F}$ controls essentially the rapidity distribution at midrapidity and at projectile and target rapidity, i.e. the baryon stopping in relativistic nucleus-nucleus collisions. We will adopt $T_{\rm F}$=0.8 fm/c for the calculations to be presented in Section 4; similar values are also used in the RQMD approach [58].

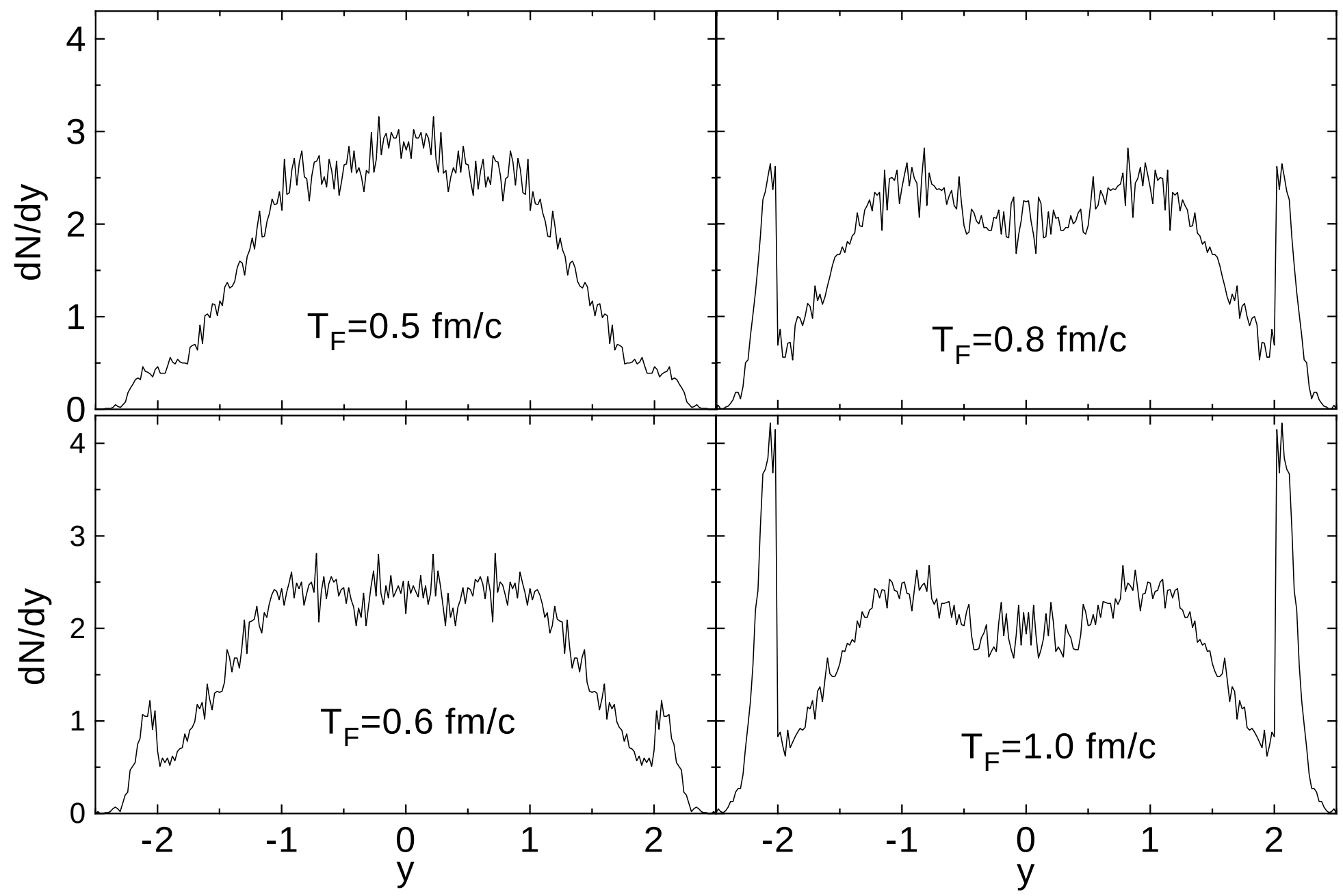

Fig. 13. The proton rapidity distribution for central collisions of $^{40}$Ca + $^{40}$Ca at 30 GeV/A using $T_{\rm F}$= 0.5, 0.6, 0.8, 1.0 fm/c, respectively.

In view of the 'chiral' dynamics addressed in this work, however, especially the productions rates of mesons should change at high baryon density [59] due to the reduced masses involved. Unfortunately, the actual meson scalar and vector selfenergies are quite a matter of debate and depend on the model parameters of the Lagrangians employed. Following the approach by Kaplan and Nelson [24] (c.f. Section 2.5), the average mass of a $K^+$ $K^-$ pair is expected to follow approximately

$$m_{\rm K^+K^-} \approx m_{\rm V}(1 - 0.16\rho_{\rm N}/\rho_0) \tag{50}$$

because the vector interaction drops out due to contributions with opposite sign and the term in $\rho_{\rm B}^2$ is rather small. Furthermore, according to the meson-baryon interaction densities (49) the in-medium mass for $\rho$, $\omega$ and $\Phi$ mesons - following Hatsuda and Lee [56]- can be approximated by

$$m(\rho_{\rm N}) \approx m_{\rm V}(1 - \lambda\rho_{\rm N}/\rho_0) \tag{51}$$

with $\lambda \approx 0.18$ for $\rho$, $\omega$ and $\lambda \approx 0.025$ for $\Phi$ mesons. The weak dependence of the $\Phi$ meson mass here is a consequence of the weak coupling of the strange quark to the light (u,d) quarks which dominantly make up the baryon density.

In view of the substantial uncertainties of the meson selfenergies especially at high density we here use this more *pragmatic model* which does not claim fundamental evidence[7]. Whereas the pion as a Goldstone boson is assumed not to change substantially with baryon density and temperature in the energy regime addressed, the kaons, K*'s, $\rho$'s, $\omega$'s and $\phi$'s are assumed to change their masses as displayed in Fig. 14 roughly in line with ref. [24,56] as pointed out above. The final values achieved at high baryon density are determined by the bare quark mass content of the mesons.

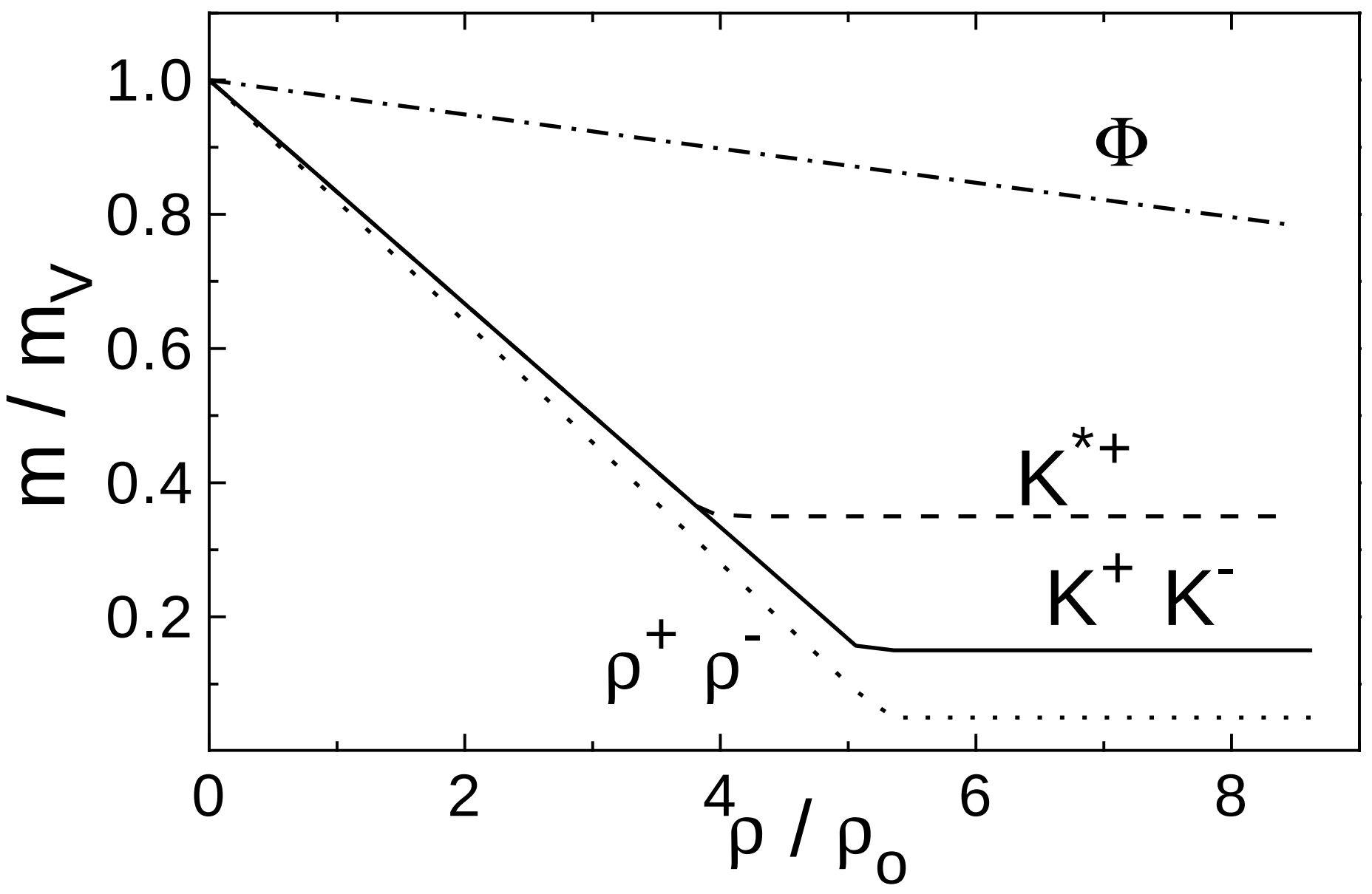

Fig. 14. Parametrization of the effective meson masses - normalized to the vacuum masses $m_V$ - versus the baryon density used in the extended string fragmentation model (HSD).

As an example for the effects to be expected at high baryon density we show in Fig. 15 the rapidity spectra of kaons and $\rho$'s for pp collisions at $T_{\rm lab}$ =20 GeV from the string fragmentation model that incorporates the density dependent meson masses from Fig. 14[8]. It is clearly seen that a dropping of the meson masses leads to a substantial enhancement of the $K^+K^-$ and $\rho$ yields and to a widening of their rapidity distribution in the individual center-of-mass system. If such phenomena can be seen in comparison to experimental data will be investigated in the next Section.

[7] These assumptions about meson properties at high baryon density can only be controlled in confrontation with sensitive experimental data.

[8] The total four-momentum is conserved in the 'chiral' string fragmentation model.

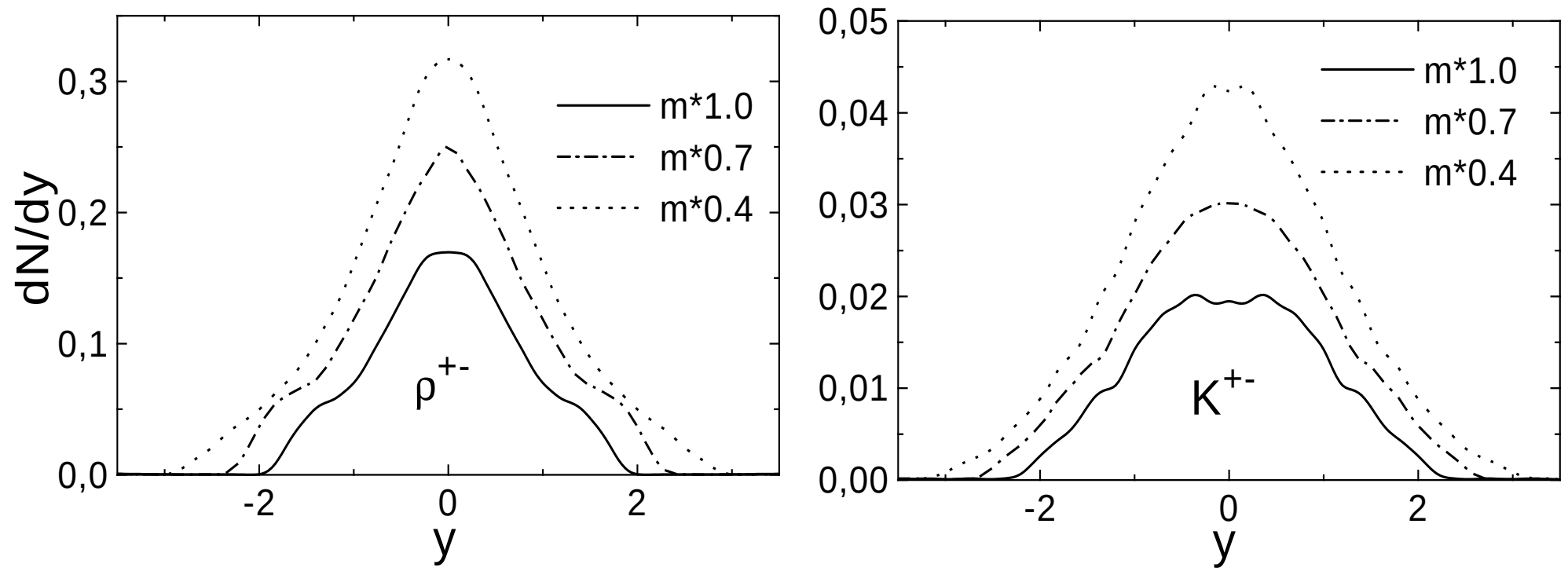

Fig. 15. Differential $K^+ + K^-$ and $\rho$ rapidity distributions from the 'chiral' string fragmentation model at 1.0, 0.7 and 0.4 $\times$ the bare masses for pp collisions at $T_{\rm lab}$=20 GeV.

## 4 Heavy-ion collisions

The relativistic transport approach (HSD) outlined in Sections 2 and 3 now is applied to nucleus-nucleus collisions from the SIS to the SPS energy regime with particular emphasis on rapidity distributions and particle spectra to control the stopping achieved in these reactions. The explicit numerical implementation of the selfenergies and collisions rates is performed in close analogy to [9,57,60,61] and does not have to be repeated here. We note that the total conservation of energy and momentum throughout the time evolution is conserved on the 2 % level for central Au + Au collisions and even better for peripheral or light-ion induced reactions.

As a first example we show in Fig. 16 the transverse $\pi^0$-spectra from Ar + Ca collisions at 1.5 GeV/A in comparison to the data of Berg et al. [62] as a characteristic system at SIS energies. Since at these energies the present approach is close to the results achieved with the former BUU model [63], the reproduction of the data is of similar quality. We thus conclude that the 'low energy dynamics' involving essentially nucleons, $\Delta$'s and pions is reasonably well included in our transport calculations.

### *4.1 Stopping in high-energy nucleus-nucleus collisions*

The next system addressed is Si + Al at 14.6 GeV/A, i.e. the AGS energy regime. The computed rapidity distribution of protons and $\pi^+$-mesons for b = 1.5 fm is compared in Fig. 17 to the data from ref. [64]. Whereas the proton rapidity distribution turns out to be quite flat in rapidity $y$ due to proton rescattering, the pion rapidity distribution is essentially of gaussian shape which reflects the pion rapidity spectrum from the string fragmentation model outlined in Section 3 (cf. Fig. 12). Similar to SIS energies [1,2] the proton rapidity distribution is insensitive to variations of the nucleon scalar and vector mean fields within the numerical accuracy.

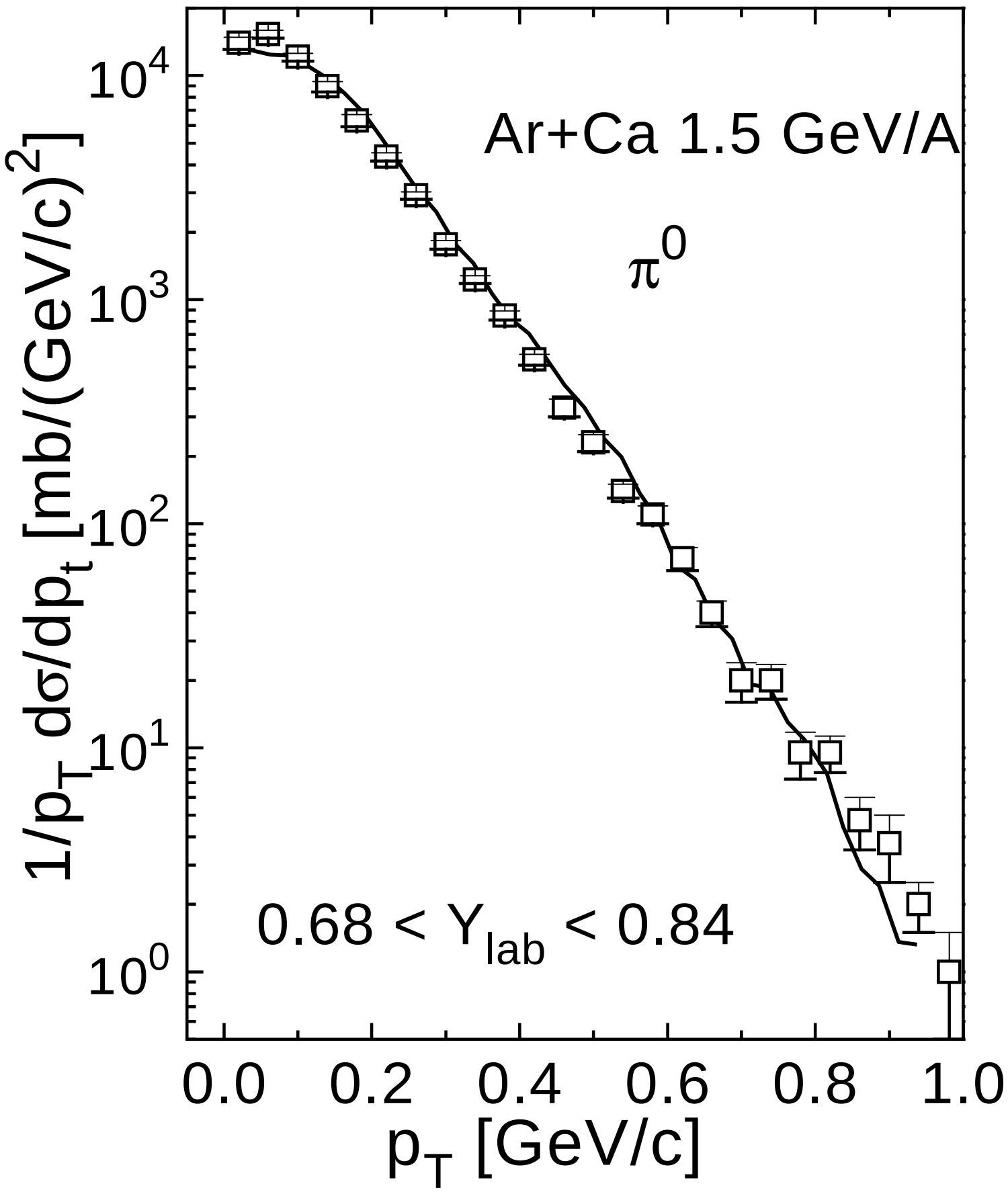

Fig. 16. Calculated transverse $\pi^0$-spectra for Ar + Ca at 1.5 GeV/A (full line) in comparison to the experimental data from ref. [62].

In analogy to Fig. 16 we show in Fig. 18 the calculated transverse mass-spectra of $\pi^+$-mesons for Si + Al at 14.6 GeV/A (solid lines) in comparison to the experimental data from ref. [64]. The overall agreement for lab. rapidities of $y = 0.9$, 1.7 and 2.7 seems to indicate that the general reaction dynamics is well reproduced within the HSD approach.

The flat proton rapidity spectrum in Fig. 17 might lead to the interpretation that there is a substantial amount of stopping in the light system Si + Al. This, however, has to be taken with care because the actual snapshots of the baryon density distribution from our computations shown in Fig. 19 (l.h.s.)

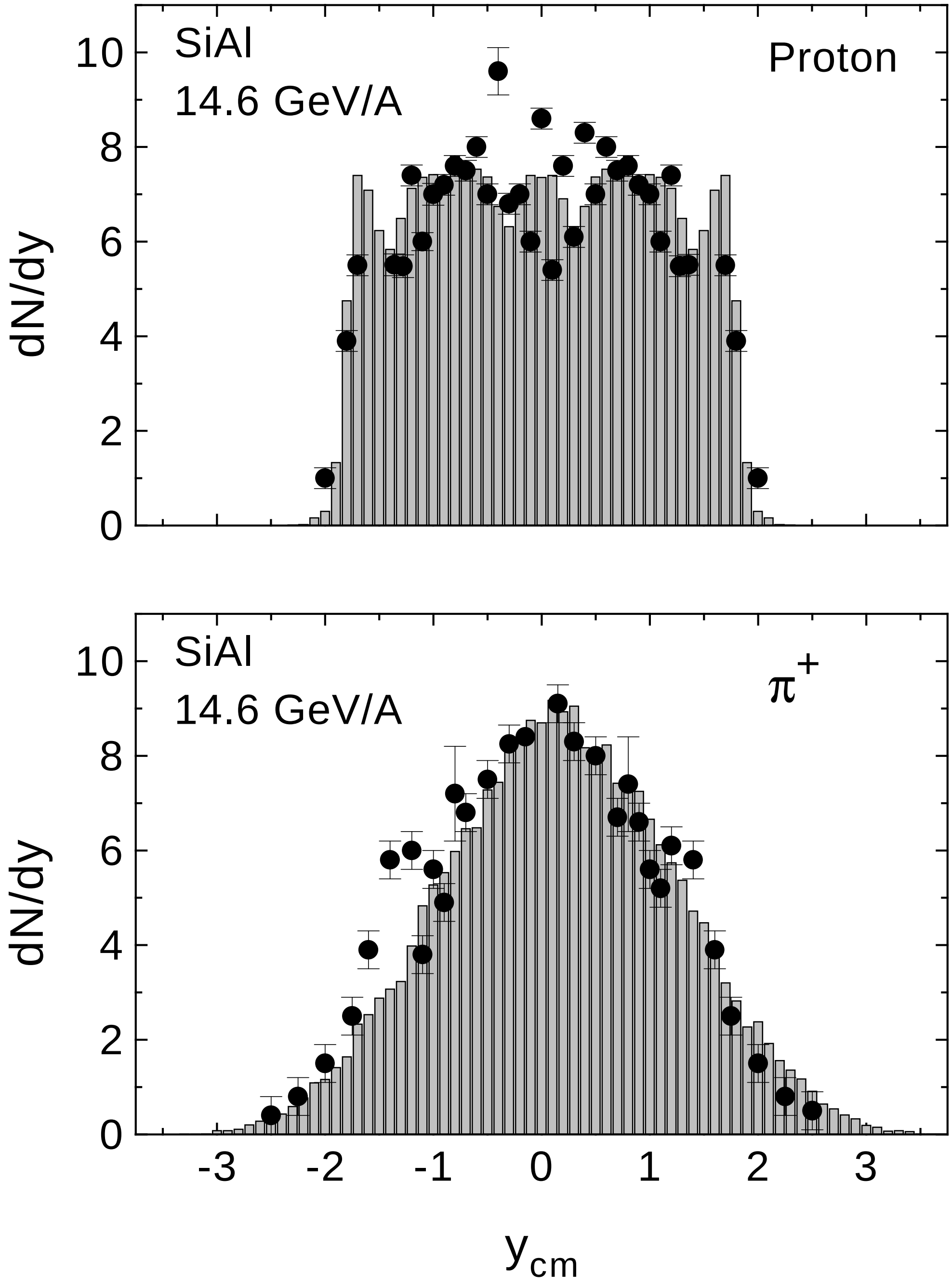

Fig. 17. Calculated proton and $\pi^+$ rapidity distribution (histrograms) for a central 14.6 GeV/A Si+Al collision in comparison to the data from [64] (full dots).

as well the phase-space distribution (r.h.s.)

$$f(z, p_z; t) = (2\pi)^{-2} \sum_b \int dr_\perp dp_\perp f_b(r_\perp, z, p_\perp, p_z; t) \ , \tag{52}$$

where $\sum_b$ denotes a sum over all baryon species, indicate a dominant transparency for the light system. This is essentially due to the large surface of the two light nuclei with a nucleon-nucleon collision probability less than 1. Furthermore, the time evolution in momentum space (middle column) shows that the system is far from kinetic equilibrium in the baryon degrees of freedom in the final state.

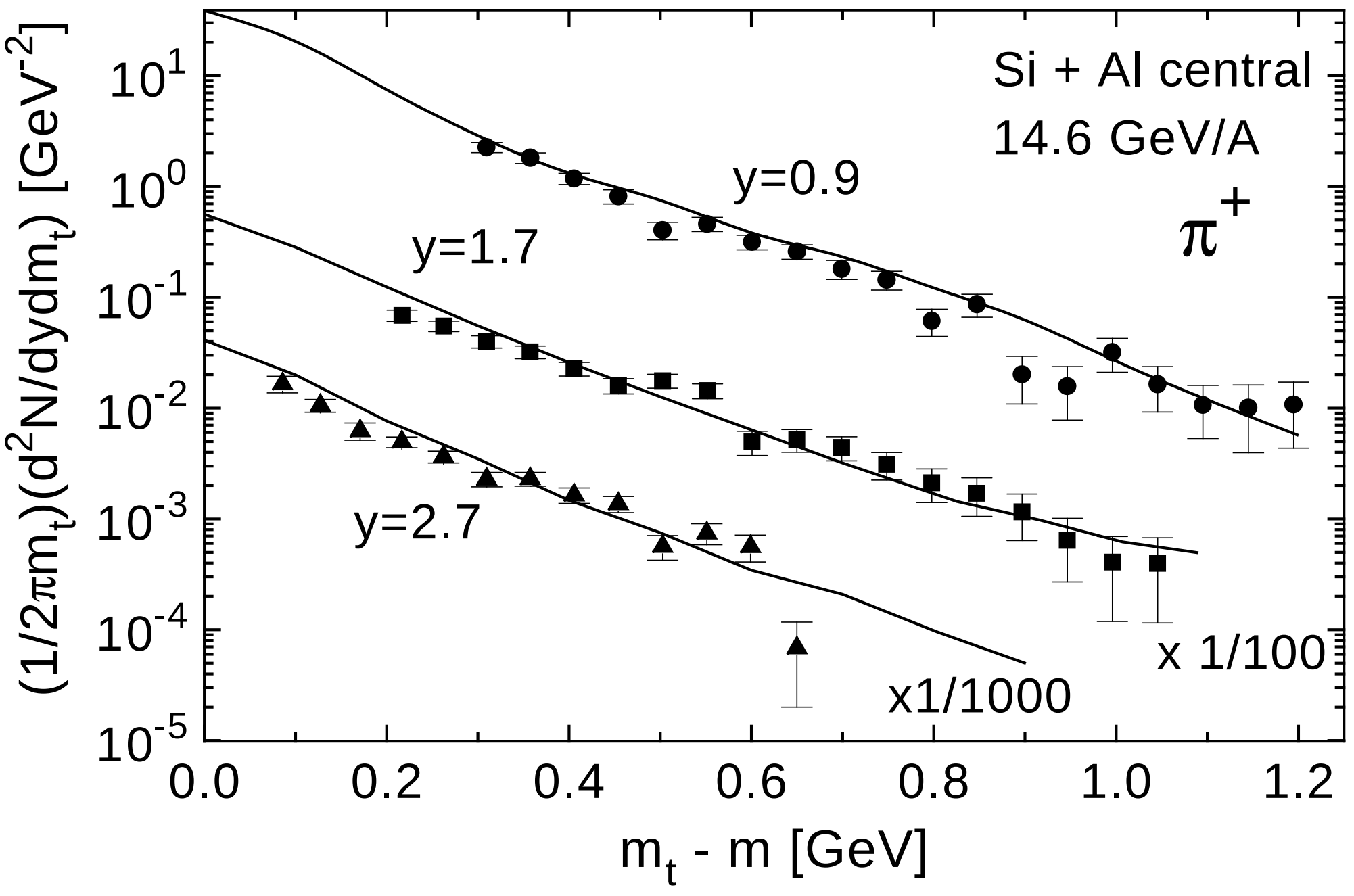

Fig. 18. Comparison of the calculated transverse mass spectra of $\pi^+$-mesons for Si + Al at 14.6 GeV/A with the experimental data from ref. [64] for rapidities $y$=0.9, 1.7, 2.7 in the laboratory system.

The amount of stopping at AGS energies is more clearly pronounced for central Au + Au reactions as displayed in Fig. 20 for the proton and $\pi^-$ rapidity distributions. Though the pion rapidity spectrum does not differ very much in shape from that of the Si + Al system in Fig. 17 at first sight, the time evolution of the baryon distribution in coordinate space, momentum space and phase space (Fig. 21) for Au+Au at 14.6 GeV/A shows a clear approach versus equilibration. However, the coordinate space evolution indicates a dominant longitudinal expansion which is also reflected in the baryon momentum distribution that does not show full isotropy. Detailed experimental data and related comparisons, however, will become available soon at the energy of 10.8 GeV/A [65]. We note that the proton rapidity spectrum for central Au + Au collisions at this energy shows a similar amount of stopping as the RQMD approach [58].

We continue our comparison to experimental data with the system S + S at 200 GeV/A, i.e. the SPS regime (Fig. 22). Though the experimental proton and $\pi^-$ rapidity spectra (from [66]) are approximately reproduced, we cannot conclude on the general applicability of our approach at SPS energies because also more simple models like HIJING or VENUS - with a less amount of rescattering - can reproduce the data in a similar way [67]. A way out of this problem is to analyze the system Pb + Pb at 153 GeV/A (Fig. 23) that has recently been studied experimentally at the SPS. Our computed proton rapidity spectrum for central collisions shows no dip at midrapidity as in HIJING or VENUS simulations [67] but a flat spectrum similar to RQMD simulations [58]. On the other hand, the pion rapidity distributions are very similar to the S + S case, however, enhanced by about a factor of 7.5.

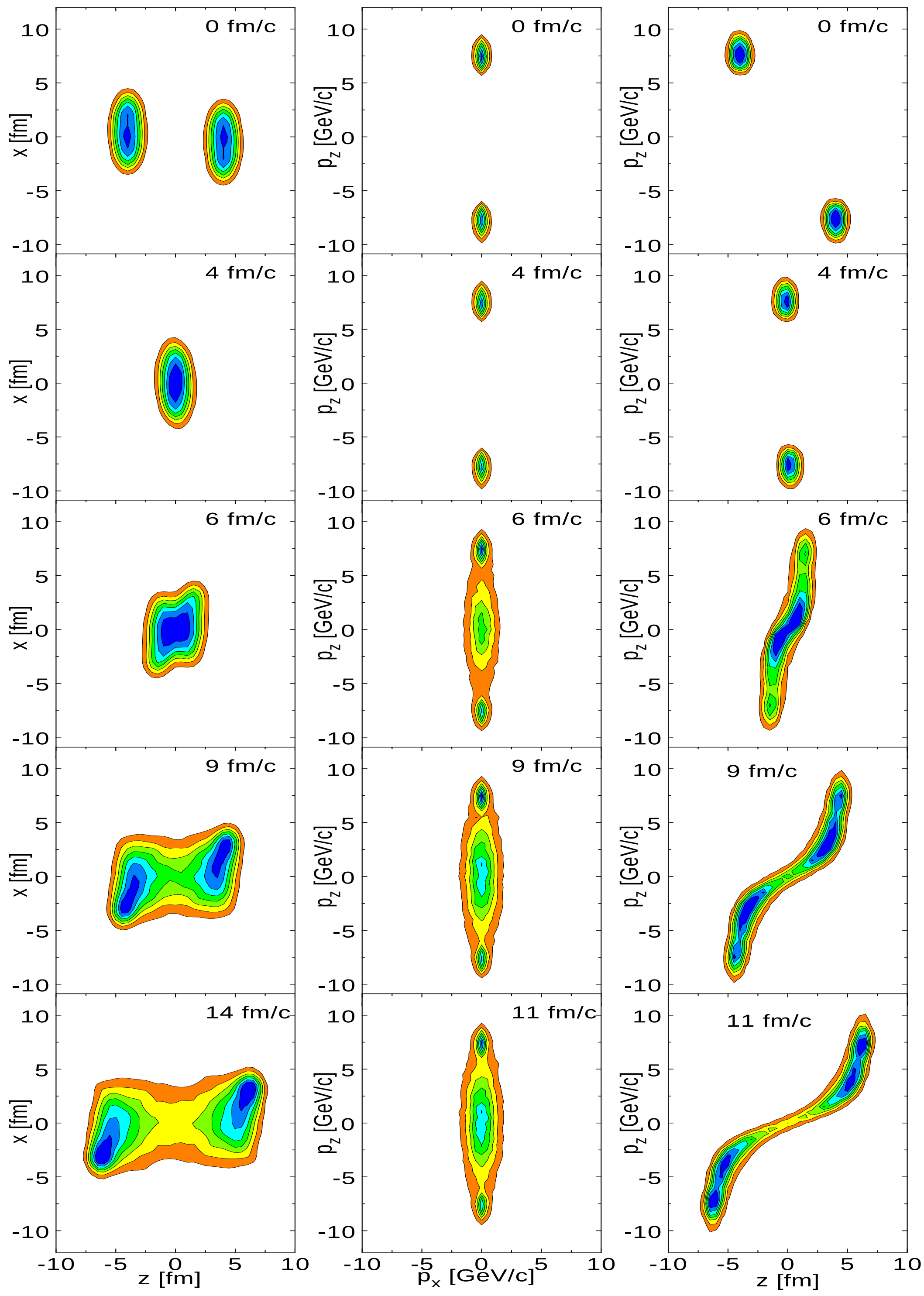

Fig. 19. Baryon density distribution (left column), momentum space (middle column) and phase-space distribution (right column) for a 14.6 GeV/A Si+Al collision at b = 1 fm for various times in fm/c.

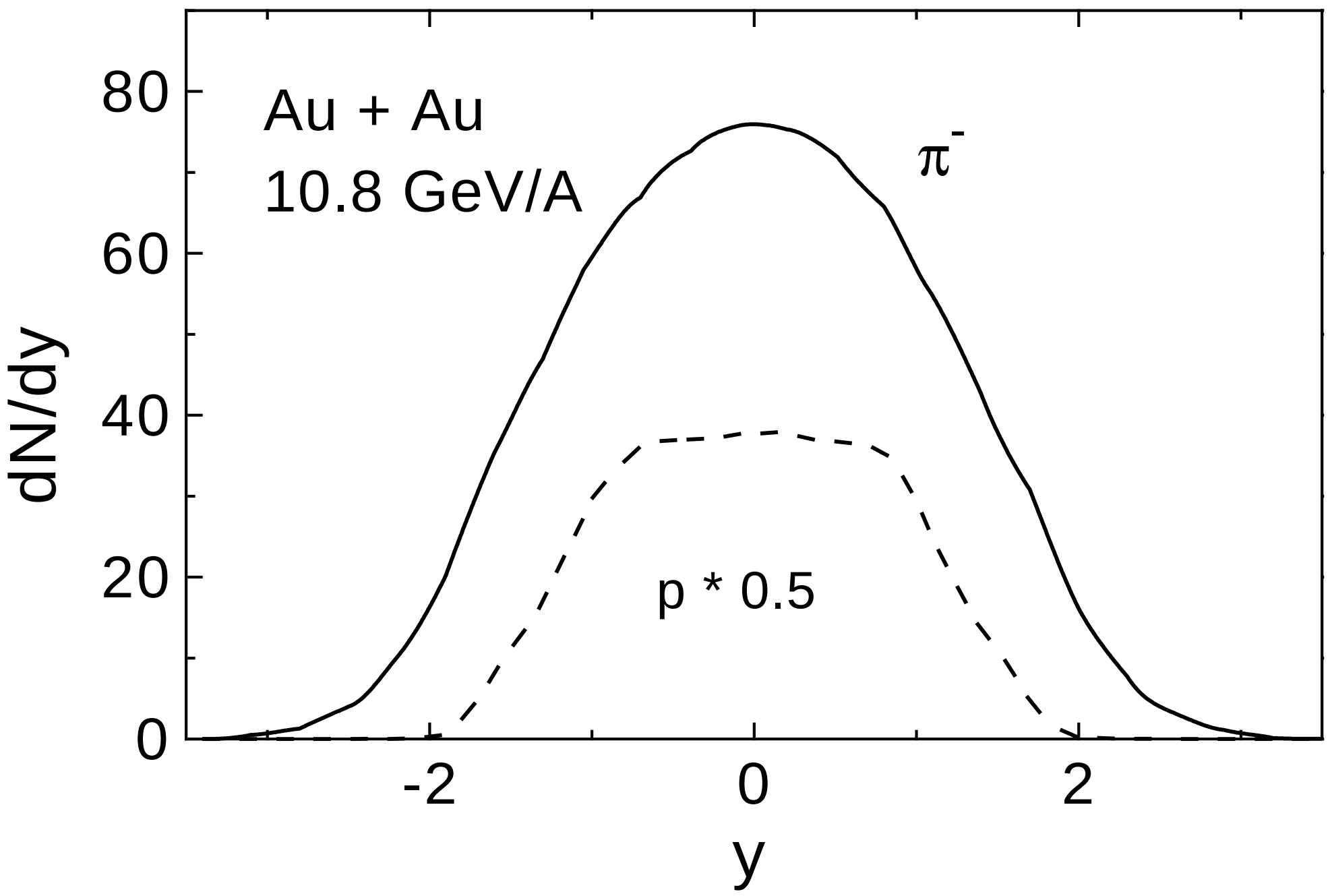

Fig. 20. Proton (dashed line) and $\pi^-$ rapidity distribution (full line) for a central 10.8 GeV/A Au + Au collision.

### *4.2 Probing chiral symmetry restoration*

The problem of chiral symmetry restoration can be investigated e.g. via the $K^+/\pi^+$ ratio as addressed in [59] since the kaon production should be enhanced due to the reduced in-medium mass (cf. Fig. 15). For this purpose we show in Table 1 the calculated $K^+/\pi^+$ yields for the systems p + p, Si + Al, Si + Au at 14.6 GeV/A and Au + Au in comparison with the experimental data for two different szenarios. The first column represents the results of a simulation where only the bare masses of the mesons have been considered in the string fragmentation approach (HSD) and all mesons are propagated as free particles whereas the second column results from density-dependent meson masses as described in Section 3. According to their drop in mass $\Delta m_{\rm h}(\mathbf{r},t) = -U_{\rm h}^{\rm S}(\mathbf{r},t)$ the mesons are propagated in their time-dependent scalar mean field $U_{\rm h}^{\rm S}(\mathbf{r},t)$ which couples linearly to the baryon density. Thus their momenta are decreased dynamically during the expansion of the hadronic system and the energy to become asymptotically on-shell is extracted from the collective motion. It is clearly seen that for density-dependent $K^+$ masses the ratio is strongly enhanced for the heavier systems as seen experimentally. However, this enhancement could also be attributed to a closer approach to chemical equilibrium as advocated in ref. [68] which might be achieved due to enhanced hadronic cross sections in the dense medium. Whereas in principle the coupled transport equations (1) also describe the approach towards chemical equilibrium for large systems, it is not yet clear if all the proper reactions rates are presently included in our simulations such that no final evidence on chiral symmetry restoration can be extracted so far from the $K^+/\pi^+$ ratio.

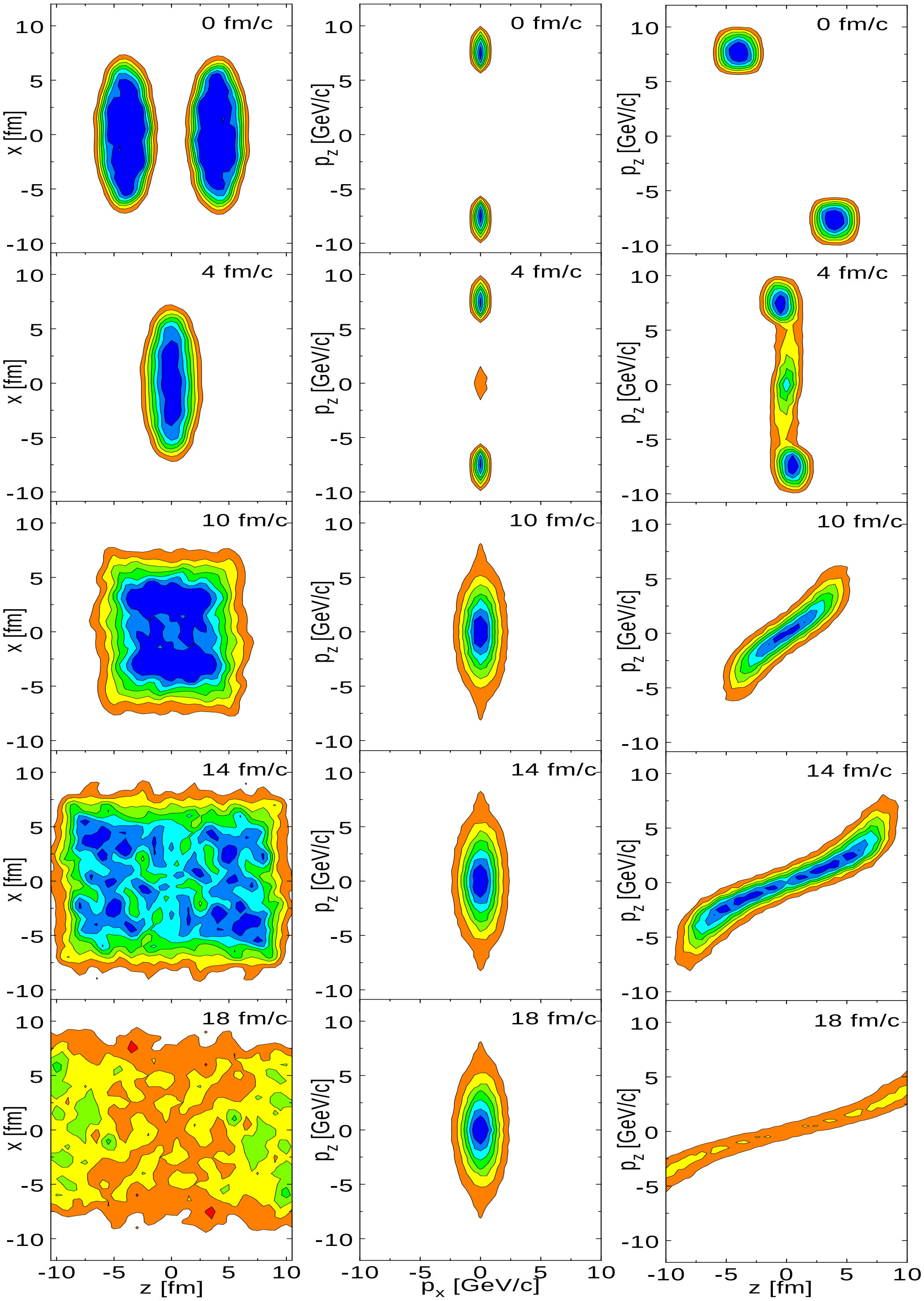

Fig. 21. Baryon density distribution (left column), momentum space (middle column) and phase-space distribution (right column) for a 14.6 GeV/A Au + Au collision at b = 0 fm for various times in fm/c.

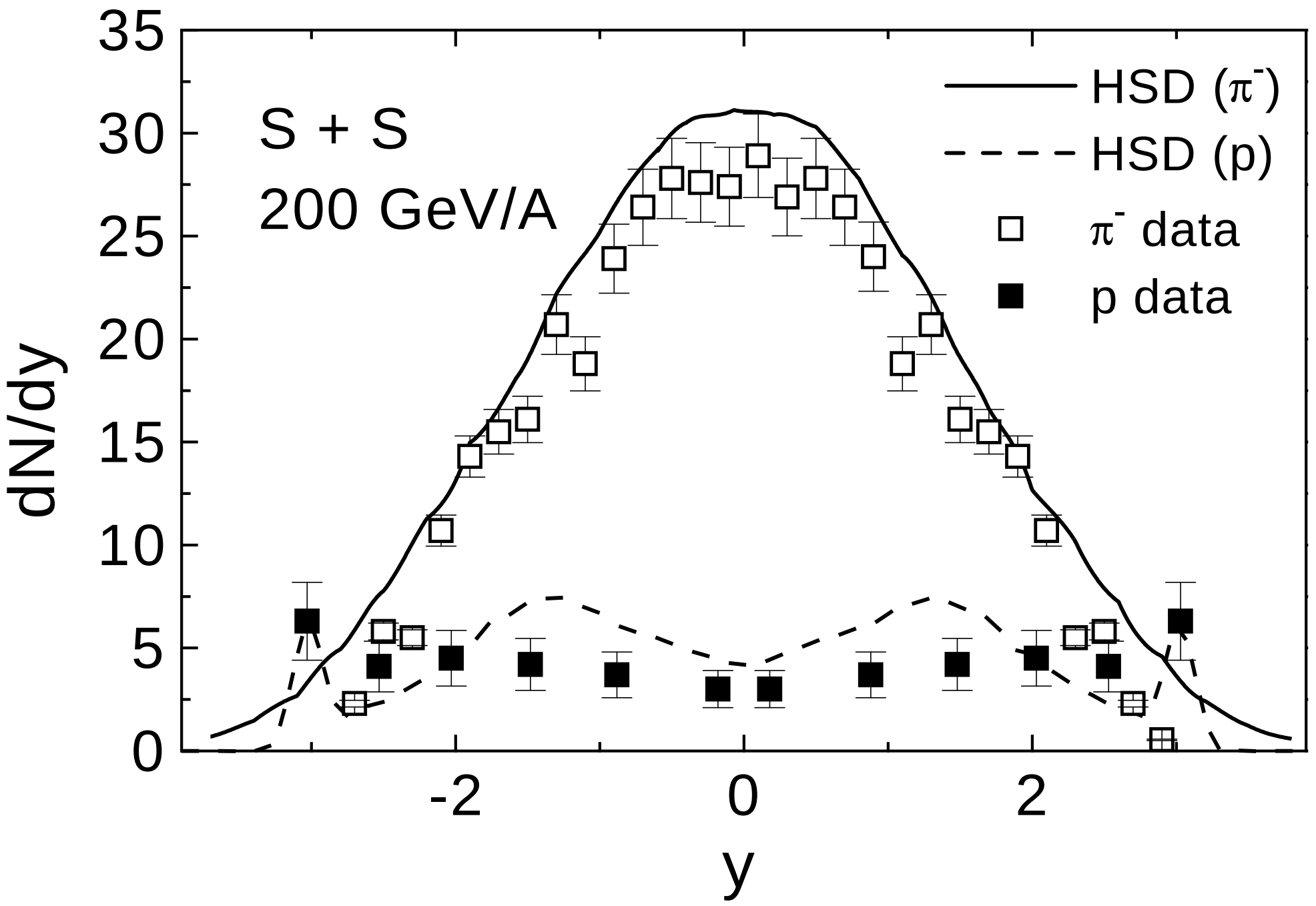

Fig. 22. Proton and $\pi^-$ rapidity distribution for a central 200 GeV/A S + S collision in comparison to the data of ref. [66].

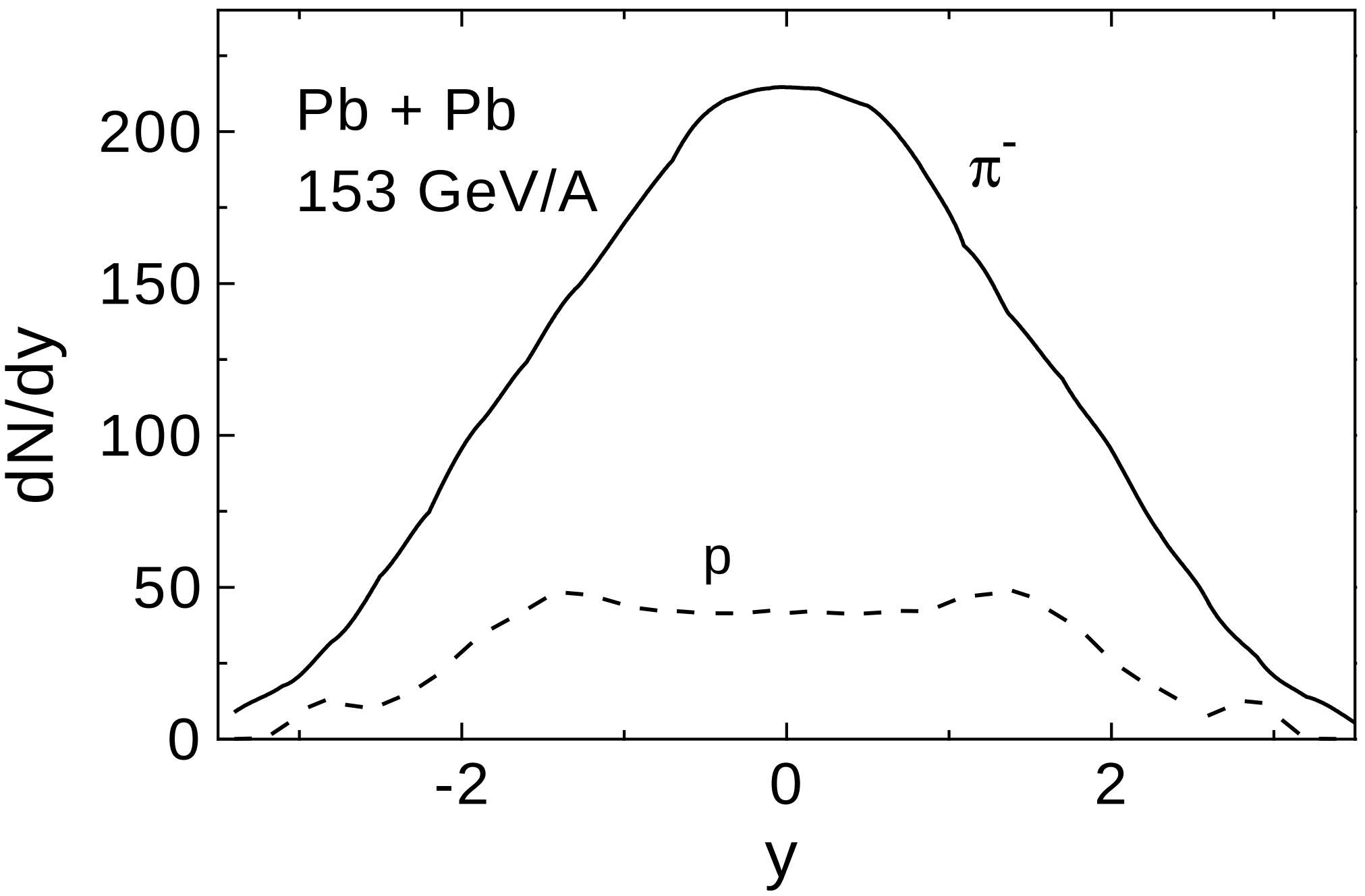

Fig. 23. Proton and $\pi^-$ rapidity distribution for a 153 GeV/A Pb + Pb collision at b = 2 fm.

| | exp.ratio | without kaon selfenergies | with kaon selfenergies |
|---|---|---|---|
| p + p | 0.08 | 0.08 | 0.08 |
| Si + Al | 0.13 | 0.09 | 0.12 |
| Si + Cu | 0.16 | 0.1 | 0.15 |
| Si + Au | 0.19 | 0.11 | 0.16 |
| Au + Au | 0.22 | 0.12 | 0.21 |

Table 1. The $K^+/\pi^+$ ratio for p + p, Si + Al, Si + Au and Au + Au collisions at 14.6 GeV/A in comparison to the data from ref. [68].

The medium modifications of the $\rho$-meson are most efficiently probed by dilepton spectroscopy [57,59,63] since due to its short lifetime the $\rho$-meson has a good chance to decay in the dense baryonic environment. According to Fig. 15 we expect a substantial enhancement of dileptons in the invariant mass range 0.4 GeV $\leq$ M $\leq$ 0.7 GeV in nucleus-nucleus collisions as compared to p + A collisions due to a shift in the $\rho$-mass spectrum and an enhanced $\rho$-meson production in the dense medium especially via $\pi^+\pi^-$ annihilation [69,70]. In fact, first computations for dilepton production within the HSD approach [70] show that the enhancement of dileptons in central S + Au collisions at 200 GeV/A (reported by the CERES-collaboration [71]) might be explained by the 'chiral' dynamics proposed in Sections 2 and 3.

## 5 Summary

In this work we have presented a relativistic transport approach for hadrons (denoted by HSD [9]) where the underlying (real parts of) nucleon selfenergies have been determined on the basis of an effective NJL-type Lagrangian for the quark degrees of freedom with a chiral invariant interaction density. Starting with a local color current interaction we have developed a model for spin and isospin averaged color neutral states on the basis of the experimental electromagnetic formfactor of the proton. The parameters in our model, which are all fixed by physical quantities are $G_{\rm S}$ and $\Lambda_{\rm S}$ for the scalar part, $G_{\rm V}$ and $\Lambda_{\rm V}$ for the vector part and $\alpha$, which describes the swelling of the nucleon in nuclear matter. The physical quantities which are sufficiently well met are: the averaged nucleon mass $M_{\rm N}$, the scalar vacuum condensate $\langle\bar{\psi}_{\rm q}\psi_{\rm q}\rangle$, the pion-nucleon $\Sigma$-term, the nuclear equation of state (minimum at $\rho = \rho_0$ with -16 MeV binding energy) and the Schroedinger equivalent potential $U_{\rm SEP}$ for nucleons. Due to the scalar-vector nature of the quark-quark interaction the nucleon selfenergies (fitted to the NJL results) are also close to those of the $\sigma$-$\omega$ model of Walecka [40] in the low momentum and low density regime.

Whereas the real part of the nucleon selfenergies has been determined from a quark oriented model to allow for reasonable extrapolations to the high density regime, the kaon, $\rho$, $\omega$ and $\phi$ meson selfenergies are fixed in line with more simple Lagrangian densities (Section 2.5). In this respect the meson sector will need further improvement in future. However, especially at bombarding energies as high as 200 GeV/A, the imaginary part of the hadron selfenergies is more important. In the HSD approach the respective transition rates have been adopted from the LUND string fragmentation model [22] where the meson masses (except the pion and $\eta$) have been scaled in line with chirally invariant interaction densities. This more pragmatic model, of cause, has less founded reliability and thus one has to justify its applicability or inadequacy in comparison to experimental data.

As a first step in this direction we have applied our relativistic transport approach to nucleus-nucleus collisions from SIS to SPS energies. Whereas the proton and pion rapidity distributions and transverse pion spectra look reasonably well for the systems studied experimentally, a clear signature for the chiral symmetry restoration could not unambiguously be established so far. This is because the strangeness enhancement observed experimentally at AGS and SPS energies might also be due to chemical equilibration or e.g. color-rope formation [58]. A better probe should be provided by dilepton spectroscopy in the invariant mass regime from 0.4 - 0.8 GeV [57] since the $\rho$-meson predominantly decays in the dense medium. In fact, first computations on $e^+e^-$ production at SPS energies suggest that the dilepton enhancement seen by the CERES collaboration might be due to a dropping $\rho$-mass in the medium [70].

[9] Hadron-String-Dynamics

The nuclear equation of state computed within our approach (Fig. 8) shows no density isomer up to $\rho \approx 10\rho_0$. This prediction is essentially due to the fact that scalar and vector baryon selfenergies are approximately of the same order of magnitude, but different sign, up to $\rho \approx 4\rho_0$ and the repulsive vector interaction takes over at even higher $\rho$ together with the kinetic energy per nucleon. A density isomer thus can only occur if the vector coupling itself decreases at high baryon density or temperature. Some arguments in this direction have recently been proposed by Brown and Rho [72] and investigated in a model study by Li and Ko [73]. A clarification of this problem e.g. should be achieved by experimental data on the baryon flow as a function of projectile/target mass and bombarding energy in the energy regime in between SIS and SPS thus allowing for a closer look at the EOS at 'very high' baryon density.

The authors acknowledge valuable and inspiring discussions througout this work with C. M. Ko, U. Mosel, H. Stöcker, S. Teis and Gy. Wolf. They are also grateful to T. Maruyama for an earlier version of the relativistic transport code used in the analysis of nucleus-nucleus collisions in the energy regime below 1 GeV/A.